\documentclass[aps,prd,twocolumn,showpacs,amsmath]{revtex4-1}

\usepackage{graphicx}

 \newcommand{\period}{{\mbox{\;.}\relax}}             
 \newcommand{\commae}{{\mbox{\;,}\relax}}             

 \newcommand{\dif}{{\rm d}}                                 
 \newcommand{\Dif}{{\rm D}}                                 
 \newcommand{\boldpartial}{\boldsymbol{\mathbf{\partial}}}  
 \newcommand{\sss}{{\,}^{\rm s\!}}                          
 \newcommand{\ci}{{\rm i}}                                  
 \newcommand{\Rea}[1]{{\,{\mathrm{Re}\,}#1}\,}              
 \newcommand{\Ima}[1]{{\,{\mathrm{Im}\,}#1}\,}              

 \newcommand{\boldu}{\mbox{\boldmath$u$}}                   
 \newcommand{\bolde}{\mbox{\boldmath$e$}}                   
 \newcommand{\boldk}{\mbox{\boldmath$k$}}                   
 \newcommand{\boldl}{\mbox{\boldmath$l$}}                   
 \newcommand{\boldm}{\mbox{\boldmath$m$}}                   
 \newcommand{\bboldm}{\bar{\mbox{\boldmath$m$}}}            
 \newcommand{\boldZ}{\mbox{\boldmath$Z$}}                   
 \newcommand{\boldV}{\mbox{\boldmath$V$}}                   

 \newcommand{\ssqrt}{{\textstyle\frac1{\sqrt{2}}}}          

\begin{document}

\title{Interpreting spacetimes of any dimension using geodesic deviation}

\author{Ji\v{r}\'{\i} Podolsk\'y}
    \email{podolsky@mbox.troja.mff.cuni.cz}
\author{Robert \v{S}varc}%
    \email{robert.svarc@mff.cuni.cz}
\affiliation{%
 Institute of Theoretical Physics, Faculty of Mathematics and Physics, Charles University in Prague,
  V Hole\v{s}ovi\v{c}k\'{a}ch 2, 180 00 Prague 8, Czech Republic
}%

\date{\today}

\begin{abstract}
We present a general method which can be used for geometrical and physical interpretation of an arbitrary spacetime in four or any higher number of dimensions. It is based on the systematic analysis of relative motion of free test particles. We demonstrate that local effect of the gravitational field on particles, as described by equation of geodesic deviation with respect to a natural orthonormal frame, can always be decomposed into a canonical set of transverse, longitudinal and Newton--Coulomb-type components, isotropic influence of a cosmological constant, and contributions arising from specific matter content of the universe. In particular, exact gravitational waves in Einstein's theory always exhibit themselves via purely transverse effects with ${D(D-3)/2}$ independent polarization states. To illustrate the utility of this approach we study the family of pp-wave spacetimes in higher dimensions and discuss specific measurable effects on a detector located in four spacetime dimensions. For example, the corresponding deformations caused by a generic higher-dimensional gravitational waves observed in such physical subspace, need not be tracefree.
\end{abstract}


\maketitle

\section{\label{sc:introduction}Introduction}

In the last decade there has been a growing interest in exact spacetimes within the context of higher-dimensional General Relativity, primarily motivated by finding particular models for string theories, AdS/CFT correspondence and brane-world cosmology. Such investigations thus concentrated mainly on various types of black holes and black rings, see \cite{EmparanReall:2008,EmparanReall:2006,Obers:2008,Kanti:2004,GibLuPagePope:2004,ChenLuPope:2006,KrtousFrolovKubiznak:2008,BertiCardosoStarinets:2009} for reviews and further references. More general static or stationary axisymmetric \cite{EmparanReall:2002,Harmark:2004,CharmousisGregory:2004,MishimaIguchi:2006,GiustoSaxena:2007,CoimbraLetelier:2007,GodazgarReall:2009}, multi-black hole Majumdar--Papapetrou-type \cite{Myers:1987,London:1995,GibbonsHorowitzTownsend:1995,Welch:1995,IvashchukMelnikov:2001,AstefaneseMannRadu:2004,CandlishReall:2007,Idaetal:2007}, and static solutions with cylindrical/toroidal symmetry \cite{HorMey99,GallowaySuryaWoolgar:2002,SarTek09b,SarTek09a,GriPod10} were also considered, including uniform and non-uniform black strings \cite{Wiseman:2003,KleihausKunzRadu:2006,CopseyHorowitz:2006,MannRaduStelea:2006,ZhaoNiuXiaDouRen:2007,BrihayeDelsate:2007,BrihayeRaduStelea:2007,BrihayeKunzRadu:2009} with the aim to elucidate their instability \cite{GregoryLaflamme:1993,HorowitzMaeda:2001,HarmarkNiarchosObers:2007}. Other important classes of higher-dimensional exact solutions of Einstein's equations have also been studied recently, for example Robinson--Trautman and Kerr--Schild spacetimes \cite{PodOrt06,OrtPodZof08,OrtPraPra09,MalekPravda:2011,GurSar02,Pod11}, extensions of the Bertotti--Robinson, (anti-)Nariai and Pleba\'{n}ski--Hacyan universes \cite{CardosoDiasLemos:2004},
higher-dimensional FLRW-type \cite{Binet:2000,ShiromizuMaedaSasaki:2000,BraxBruckDavis:2004,GarciaCarlip:2007,MaartensKoyama:2010} and multidimensional cosmological models \cite{IvashchukMelnikov:2001,Hervik:2002} (see also references therein), specific solitons \cite{HorMey99,ClarksonMann:2006a,ClarksonMann:2006b}, or various exact gravitational waves --- in particular those which belong to non-expanding Kundt family \cite{PodZof09,ColHerPapPel09}, namely generalized pp-waves \cite{Bri25,GibbonsRuback:1986,ColMilPelPraPraZal03,ColeyHervik:2004,Ort04,ColMilPraPra04,ColFusHerPel06} (for a study of their collisions see \cite{GursesKarasu:2001}), VSI \cite{ColMilPraPra04,ColFusHerPel06} and CSI \cite{ColHerPel06} spacetimes, or relativistic gyratons \cite{FroFur05,FroIsrZel05,FroZel05,FroZel06,FroLin06,CalLeZor07}.

Fundamental general questions concerning the classification of higher-dimensional manifolds based on the algebraic structure of the curvature tensor have been clarified \cite{Coley:2008,ColeyMilsonPravdaPravdova:2004,Ort09,Sen10}, including generalizations of the Newman--Penrose and the Geroch--Held--Penrose formalisms \cite{PravdaPravdovaColeyMilson:2004,OrtPraPra07,PraPra08,DurPraPraReall10,DurkeeReall:2011}. This paved the way for a systematic study of wide classes of algebraically special spacetimes in higher dimensions \cite{PraPraOrt07,Durkee:2009b,OrtPraPra10,OrtPraPra11}. Investigation of asymptotic behaviour of the corresponding fields and their global structure, in particular properties of gravitational radiation, has also been initiated \cite{MarolfRoss:2002,CarDiaLemos03,HollandsWald:2004,HollandsIshibashi:2005,Ishibashi:2008,PravdovaPravdaColey:2005,AlesciMontani:2005,KrtousPodolsky:2004b,KrtousPodolsky:2006,AndersonChrusciel:2005,BizonChmajSchmidt:2005,ChoquetChruscielLoizelet:2006,BeigChrusciel:2007,MironovMorozon:2008,OrtaggioPravdaPravdova:2009}.

Nevertheless, in spite of the considerable effort devoted to this topic, there are still important aspects concerning the nature of gravitational fields in higher-dimensional gravity that remain open. Any sufficiently general method which could be used to probe geometrical and physical properties of a given spacetime would be useful. In the present work we suggest and develop such an approach which is based on investigation and classification of specific effects of gravity encoded in relative motion of nearby test particles.

In fact, in standard four-dimensional General Relativity, this has long been used as an important tool for studies of spacetimes. Relative motion of close free particles  helps us to clarify the structure of a gravitational field in which the test particles move. When they have no charge and spin, this is mathematically described by the {\em equation of geodesic deviation} (sometimes also called the Jacobi equation) which was first derived in the $n$-dimensional (pseudo-)Riemannian geometry by Levi-Civita and Synge \cite{LeviCivita:1926,Synge:1926,Synge:1934,SyngeSchild:book1949}, see \cite{Trautman:2009} for the historical account. Shortly after its application to Einstein's gravity theory \cite{Pirani:1956,Pirani:1957,BondiPiraniRobinson:1959,Sachs:1960,Weber:1960,Synge:book1960,Weber:book1961,EhlersKundt:1962,Pirani:1965,Szekeres:1965} it helped, for instance, to understand the behaviour of test bodies influenced by gravitational waves or the physical fate of observers falling into black holes. Textbook descriptions of this equation, which is linear with respect to the separation vector connecting the test particles, are given, e.g., in \cite{MisnerThorneWheeler:book,Wald:book,Hobsonetal:book,FeliceBini:book}. Let us also mention that generalizations of the equation of geodesic deviation to admit arbitrary relative velocities of the particles were obtained in the works \cite{Hodgkinson:1972,Mashhoon:1975,Mashhoon:1977,Bazanski:1977,AlexandrovPiragas:1979,LiNi:1979,AudretschLammerzahl:1983,Ciufolini:1986,CiufoliniDemianski:1986,ChiconeMashhoon:2002}. Further extensions, higher-order corrections to the geodesic deviation equation, their particular applications and references can be found in the recent papers \cite{ChiconeMashhoon:2002,BicakPodolsky:1999b,BalakinHoltenKerner:2000,Manoff:2001,KernerHoltenColistete:2001,ColisteteLeygnacKerner:2002,ChiconeMashhoon:2006,MullariTammelo:2006} and in the monograph \cite{FeliceBini:book}.

In 1965 Szekeres \cite{Szekeres:1965} presented an elegant analysis of the behaviour of nearby test particles in a \emph{generic} four-dimensional spacetime. He demonstrated that the overall effect consists of specific transverse, longitudinal and Newton--Coulomb-type components. This was achieved by  decomposing the Riemann curvature tensor into the Weyl tensor and the terms involving the Ricci tensor (and Ricci scalar). While the former represents the ``free gravitational field'' the latter can be explicitly expressed, employing Einstein's field equations, in terms of the corresponding components of the energy-momentum tensor which describes the matter content. In order to further analyze the Weyl tensor contribution, Szekeres used the formalism of self-dual bivectors \cite{JordanEhlersKundt:1960,Sachs:1961} constructed from {\em null frames}. This enabled him to deduce the effects of gravitational fields on nearby test particles in spacetimes of various Petrov types. When these results are re-expressed in a more convenient Newman--Penrose formalism \cite{NewmanPenrose:1962,PenroseRindler:book}, explicit physical interpretation of the corresponding complex scalars $\Psi_A$ are obtained. In particular, the Weyl scalar $\Psi_4$ (the only non-trivial component in type~N spacetimes) represents purely transverse effect of exact gravitational waves, the scalar $\Psi_3$ (present, e.g., in type~III spacetimes) is responsible for longitudinal effects, and $\Psi_2$ (typical for spacetimes of type~D) gives rise to Newton-like deformations of the family of test particles  (see \cite{Griffiths:book,Stephanietal:book,GriffithPodolsky:book} for more details; inclusion of a nonvanishing cosmological constant was described in \cite{BicakPodolsky:1999b}).

It is the purpose of the present work to extend these results to arbitrary spacetimes in {\em any} dimension ${D\ge4}$. The paper is organized as follows. In section~2 we recall the equation of geodesic deviation, including its invariant form with respect to the interpretation orthonormal frame adapted to an observer. In section~3 we perform the canonical decomposition of the curvature tensor using Einstein's equations and the real Weyl tensor components $\Psi_{A^{...}}$ with respect to an associated null frame. We thus derive an explicit and general form of the equation of geodesic deviation. Section~4 analyses the character of all canonical components of a gravitational field. Section~5 is devoted to the discussion of  uniqueness of the interpretation frame, and derivation of explicit relations which give the dependence of the field components on the observer's velocity. In section~6 we describe the effect of pure radiation, perfect fluid and electromagnetic field on test particles. Final section~7 illustrates the method on the family of pp-waves in higher dimensions. There are also 3 appendices: In appendix~A we give relations to the standard complex formalism of ${D=4}$ General Relativity, and in appendix~B we summarize alternative notations commonly used in literature on ${D\ge4}$ spacetimes. Finally, in appendix~C the Lorentz transformations of the $\Psi_{A^{...}}$ scalars are presented.

\section{\label{sc:geodev}Equation of geodesic deviation}

The main objective of the present work is to investigate and characterize the curvature of an arbitrary spacetime of dimension ${D \ge 4}$ by its local effects on freely falling test particles (observers). The gravitational field manifests itself, in Newtonian terminology, as specific ``tidal forces'' which cause the nearby particles to accelerate relative to each other. This leads to a deviation of corresponding geodesics whose separation thus changes with time: in various spatial directions the particles approach or recede from themselves, exhibiting thus the specific character of the spacetime in the vicinity of a given event.

\begin{figure}[ht]
   \includegraphics[scale=0.70]{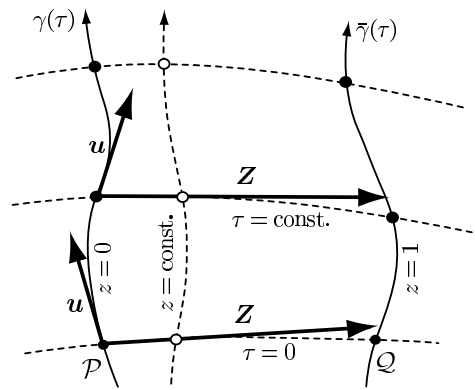}%
   \caption{\label{fig1}%
   In a curved $D$-dimensional spacetime, nearby test particles moving along geodesics accelerate toward or away from each other, as given by the equation of geodesic deviation \eqref{EqGeoDev}. Here $\boldu$ is the velocity vector of a reference particle, and $\boldZ$ is the separation vector which represents actual relative position of the second test particle at a given proper time $\tau$.
   }
   \end{figure}

In standard and also higher-dimensional General Relativity, such a behaviour of free test particles (without charge and spin) is described by the geodesic deviation equation
\cite{LeviCivita:1926,Synge:1926,Synge:1934,SyngeSchild:book1949,Trautman:2009,Pirani:1956,Pirani:1957,BondiPiraniRobinson:1959,Sachs:1960,Weber:1960,Synge:book1960,Weber:book1961,EhlersKundt:1962,Pirani:1965,Szekeres:1965,MisnerThorneWheeler:book,Wald:book,Hobsonetal:book,FeliceBini:book}
\begin{equation}\label{EqGeoDev}
\frac{\Dif^2 Z^\mu}{\dif\,\tau^2}=R^\mu_{\ \alpha\beta\nu}\,u^\alpha u^\beta Z^\nu \commae
\end{equation}
where $R^\mu_{\ \alpha\beta\nu}$ are components of the \emph{Riemann curvature tensor}, $u^\alpha$ are components of the \emph{velocity vector} ${\boldu=u^\alpha\boldpartial_\alpha}$ of the reference (fiducial) particle moving along a timelike geodesic ${\gamma(\tau) \equiv\{x^0(\tau), \ldots, x^{D-1}(\tau)\}}$,  ${u^\alpha=\frac{\dif x^\alpha}{\dif\tau}}$, the parameter $\tau$ is its proper time (so that ${\boldu\cdot\boldu\equiv g_{\alpha\beta}\,u^\alpha u^\beta = -1}$), and $Z^\mu$ are components of the \emph{separation vector} ${\boldZ=Z^\mu\boldpartial_\mu}$ which connects the reference particle with another nearby test particle moving along a timelike geodesic ${\bar{\gamma}(\tau)}$. The situation is visualized in figure~\ref{fig1}.

Equation \eqref{EqGeoDev} explicitly expresses the relative acceleration of two nearby particles by the second absolute (covariant) derivative of the vector field $\boldZ$ along $\gamma(\tau)$,
\begin{equation}\label{CovarDer}
\frac{\Dif^2 Z^\mu}{\dif\,\tau^2} = \left(Z^\mu_{\ ;\gamma} u^\gamma\right)_{;\delta} u^\delta = Z^\mu_{\ ;\gamma\delta}\, u^\gamma u^\delta \commae
\end{equation}
in terms of the local curvature tensor and the actual relative position of the particles, described by the separation vector $\boldZ(\tau)$ at the time $\tau$.

To be geometrically more precise, the two geodesics should be understood as specific representatives of a congruence $\gamma(\tau,z)$, i.e. smooth one-parameter family of geodesics, such that ${\gamma(\tau)\equiv\gamma(\tau,z=0)}$ and ${\bar{\gamma}(\tau)\equiv\gamma(\tau,z=\hbox{const.})}$. The proper time~$\tau$ and the parameter $z$ which labels the geodesics can be chosen as coordinates on the submanifold spanned by the congruence. Thus ${\boldu=\boldpartial_\tau}$ and ${\boldZ=\boldpartial_z}$, and the deviation vector field $\boldZ$ is Lie-transported along the geodesics generated by $\boldu$. Consider now the positions of two test particles at a given time, for example ${\cal P}$ (located at ${z=0}$) and ${\cal Q}$ (for which ${z=1}$, say) at ${\tau=0}$, as shown in figure~\ref{fig1}. Their coordinates are related by the exponential map ${x^\mu_{\cal Q} = \exp(z\boldZ)\,x^\mu_{\cal P}}$ generated by ${\boldZ}$ at ${\cal P}$, where we set ${z=1}$ to locate ${\cal Q}$. If the higher-order terms are negligible this expression reduces to ${x^\mu_{\cal Q}-x^\mu_{\cal P}\approx (\boldZ\,x^\mu)_{\cal P}}$, demonstrating that the separation vector $\boldZ$ describes the relative position of the two test particles, and $\boldZ(\tau)$ gives its evolution that is obtained by solving the equation \eqref{EqGeoDev}. Such linear approximation improves when the second test particle moves very close to the reference one, i.e. along the geodesic ${z=\hbox{const.}\ll 1}$, in which case the separation is described by the vector field $z\boldZ(\tau)$.

It should also be recalled that the equation of geodesic deviation \eqref{EqGeoDev} is linear with respect to the components of the separation vector, neglecting higher-order terms in the Taylor expansion of exact expression for relative acceleration of free test particles. It can thus be used when the relative velocities of the particles are negligible, i.e. their geodesics are almost parallel. Generalizations of equation \eqref{EqGeoDev} to admit arbitrary relative velocities were obtained and applied in the works \cite{Hodgkinson:1972,Mashhoon:1975,Mashhoon:1977,Bazanski:1977,AlexandrovPiragas:1979,LiNi:1979,Ciufolini:1986,CiufoliniDemianski:1986,ChiconeMashhoon:2002}. Further extensions, higher-order corrections to the geodesic deviation equation and their specific applications can be found in \cite{BalakinHoltenKerner:2000,Manoff:2001,KernerHoltenColistete:2001,ColisteteLeygnacKerner:2002,ChiconeMashhoon:2006,MullariTammelo:2006} (for reviews and other references see \cite{FeliceBini:book,ChiconeMashhoon:2002,Manoff:2001,MullariTammelo:2006}). Our aim, however, in this paper is to investigate local relative motion of nearby free test particles which are initially at rest with respect to each other. For such an analysis, the classical geodesic deviation equation \eqref{EqGeoDev} will be fully sufficient.

Now, in order to obtain \emph{invariant results} independent of the choice of coordinates, it is natural to adopt the Pirani approach \cite{Pirani:1956,Pirani:1957} based on the use of components of the above quantities with respect to a suitable \emph{orthonormal frame} ${\{\bolde_a \}}$. At any point of the reference geodesic this defines an observer's framework in which physical measurements are made and interpreted. In particular, the separation vector is expressed as ${\boldZ=Z^a \,\bolde_a}$. The timelike vector of the frame is identified with the velocity vector of the observer, ${\bolde_{(0)}=\boldu}$, and ${\bolde_{(\rm{i})}}$, where ${{\rm{i}}=1,2,\ldots, D-1}$, are perpendicular spacelike unit vectors which form its local Cartesian basis in the hypersurface orthogonal to $\boldu$ (see also figure~\ref{fig2}),
\begin{equation}\label{DefFrame}
\bolde_a \cdot \bolde_b \equiv g_{\alpha\beta}\, e_a^\alpha e_b^\beta=\eta_{ab}\equiv\hbox{diag}(-1,1,\ldots,1) \period
\end{equation}
Due to the fact that $\boldu$ is parallelly transported, for the zeroth frame-component ${Z^{(0)}\equiv\bolde^{(0)}\cdot\boldZ=-\boldu\cdot\boldZ}$ we immediately obtain
\begin{equation}\label{ZeroComp}
\frac{\dif^2Z^{(0)}}{\dif\,\tau^2}=-u_\mu \frac{\Dif^2Z^\mu}{\dif\,\tau^2} = -R_{\mu\alpha\beta\nu}\,u^\mu u^\alpha u^\beta Z^\nu =0  \commae
\end{equation}
using the skew-symmetry of the Riemann tensor. Therefore, ${Z^{(0)}(\tau)}$ must be at most a linear function of the proper time. By a natural choice of initial conditions, consistent with the above construction of the geodesic congruence $\gamma(\tau,z)$, we set ${Z^{(0)}=0}$. The temporal component of $\boldZ$ thus vanishes and the test particles always stay in the same spacelike hypersurfaces synchronized by~$\tau$.

Physical information about relative motion of the test particles is thus completely contained in the spatial frame components ${Z^{(\rm{i})}(\tau)\equiv\bolde^{(\rm{i})}\cdot\boldZ}$ of the separation vector~$\boldZ$. These determine the actual \emph{relative spatial position} of the two nearby particles. By projecting the geodesic deviation equation \eqref{EqGeoDev} onto ${\bolde^{(\rm{i})}=\bolde_{(\rm{i})}}$ we obtain
\begin{equation}\label{InvGeoDev}
\ddot Z^{(\rm{i})}= R^{(\rm{i})}_{\quad(0)(0)(\rm{j})}\,Z^{(\rm{j})} \commae
\end{equation}
where ${\rm{i},\rm{j}=1,2,\ldots, D-1}$, and we denote the \emph{physical relative acceleration} as
\begin{equation}\label{PhysAccel}
\ddot Z^{(\rm{i})} \equiv \bolde^{(\rm{i})}\cdot \frac{\Dif^2\boldZ}{\dif\, \tau^2}=e^{(\rm{i})}_\mu\,\frac{\Dif^2Z^\mu}{\dif\, \tau^2} \period
\end{equation}
The frame components of the Riemann tensor are ${R_{(\rm{i})(0)(0)(\rm{j})}\equiv R_{\mu\alpha\beta\nu} \,e^\mu_{(\rm{i})}u^\alpha u^\beta e^\nu_{(\rm{j})}}$.
Let us note that Pirani \cite{Pirani:1956,Pirani:1957} labeled, in ${D=4}$, the frame components of the ``tidal stress tensor'' that occurs in equation \eqref{InvGeoDev} (with an opposite sign) as ${{K^a}_b\equiv{R^a}_{0b0} = {R^a}_{cbd}\,u^cu^d}$. They are equivalent to the \emph{electric part} of the Riemann tensor ${{\cal E}_{ab}\equiv R_{a0b0} = R_{acbd}\,u^cu^d}$, see \cite{FeliceBini:book}.

Following Pirani it is also usually assumed that the orthonormal frame ${\{\bolde_a \}}$ is parallelly propagated along the reference geodesic. However, in our work we do not make such an assumption. In fact, as a key idea of the proposed interpretation method, we align the orthonormal frame with the \emph{algebraic structure} of a given spacetime instead (see also section~\ref{sc:framechoice}). This makes the investigation of its physical properties much easier.

\section{\label{sc:canondecomp}Canonical decomposition of the curvature tensor}

Next step is to express the frame components of the Riemann tensor ${R_{(\rm{i})(0)(0)(\rm{j})}}$. Using the standard decomposition of the curvature tensor into the traceless Weyl tensor $C_{abcd}$ and specific combinations of the Ricci tensor $R_{ab}$ and Ricci scalar ${R\,}$,
\begin{eqnarray}\label{Decomp}
R_{abcd}&=&C_{abcd}+\frac{2}{D-2}\left(\,g_{a[c}\,R_{d]b}-g_{b[c}\,R_{d]a}\right) \commae \nonumber\\
  && -\frac{2}{(D-1)(D-2)}\,R\,g_{a[c}\,g_{d]b}
\end{eqnarray}
we immediately obtain
\begin{eqnarray}\label{DecompFrame}
R_{(\rm{i})(0)(0)(\rm{j})}&=&C_{(\rm{i})(0)(0)(\rm{j})}+\frac{1}{D-2}\Big(R_{(\rm{i})(\rm{j})}-\delta_{\rm{i}\rm{j}}\,R_{(0)(0)}\Big) \nonumber\\
  && -\frac{R\,\delta_{\rm{i}\rm{j}} }{(D-1)(D-2)} \period
\end{eqnarray}
Before substituting this into the geodesic deviation equation \eqref{InvGeoDev} we also employ the Einstein field equations, generalized to any dimension ${D\ge4\,}$,
\begin{equation}\label{Einstein}
R_{ab}-{\textstyle\frac{1}{2}}R\, g_{ab} + \Lambda\,g_{ab}=8\pi\,T_{ab}  \commae
\end{equation}
where $\Lambda$ is a \emph{cosmological constant} and ${T_{ab}}$ is the \emph{energy-momentum tensor} of the matter field. Using \eqref{Einstein} and its trace ${R=\frac{2}{2-D}(8\pi\,T-D\,\Lambda)}$, we rewrite \eqref{DecompFrame} as
\begin{eqnarray}
R_{(\rm{i})(0)(0)(\rm{j})} &=& \frac{2\Lambda\,\delta_{\rm{i}\rm{j}}}{(D-1)(D-2)}+C_{(\rm{i})(0)(0)(\rm{j})}\\
 && +\frac{8\pi}{D-2}\left[T_{(\rm{i})(\rm{j})}-\delta_{\rm{i}\rm{j}}\Big(T_{(0)(0)}+\frac{2\,T}{D-1}\Big)\right]. \nonumber
\end{eqnarray}
The equation of geodesic deviation \eqref{InvGeoDev} thus takes the following invariant form
\begin{eqnarray}\label{InvGeoDevExpl}
\ddot{Z}^{(\rm{i})}&=&  \frac{2\Lambda }{(D-1)(D-2)}\,Z^{(\rm{i})} + C_{(\rm{i})(0)(0)(\rm{j})}\,Z^{(\rm{j})}\\
               &&  +\frac{8\pi}{D-2}\left[\,T_{(\rm{i})(\rm{j})} \,Z^{(\rm{j})}-\Big(T_{(0)(0)}+\frac{2}{D-1}\,T\Big)\, Z^{(\rm{i})}\,\right]. \nonumber
\end{eqnarray}
The first term represents the isotropic influence of the cosmological constant $\Lambda$ on free test particles, the second term describes the effect of a ``free'' gravitational field encoded in the Weyl tensor, while the second line in \eqref{InvGeoDevExpl} gives a direct effect of specific matter present in a given spacetime.

The terms proportional to the coefficients ${C_{(\rm{i})(0)(0)(\rm{j})}}$ can further be conveniently expressed using the Newman--Penrose-type scalars, which are the components of the Weyl tensor with respect to an associated (real) \emph{null frame}  ${\{\boldk, \boldl, \boldm_{i} \}}$. This frame is introduced by the relations
\begin{eqnarray}\label{NullFrame}
 && \boldk=\ssqrt(\boldu+\bolde_{(1)})\,, \quad \boldl=\ssqrt(\boldu-\bolde_{(1)})\,, \nonumber\\
 && \boldm_{i}=\bolde_{(i)} \hspace{2mm} \hbox{for} \hspace{2mm} i=2,\ldots,D-1\commae
\end{eqnarray}
where ${\boldu\equiv\bolde_{(0)}}$ is the velocity vector of the observer. Thus $\boldk$ and $\boldl$ are future oriented null vectors, and $\boldm_{i}$ are ${D-2}$ spatial Cartesian vectors orthogonal to them, satisfying
\begin{eqnarray}\label{NullFrNorm}
 && \boldk\cdot\boldl=-1\,, \quad \boldm_{i}\cdot\boldm_{j}=\delta_{ij}\,, \nonumber\\
 && \boldk\cdot\boldk=0=\boldl\cdot\boldl\,, \quad \boldk\cdot\boldm_{i}=0=\boldl\cdot\boldm_{i}\period
\end{eqnarray}

Using the notation of \cite{KrtousPodolsky:2006}, the components of the Weyl tensor in such a null frame are determined by the following scalars (grouped by their boost weight):
\begin{eqnarray}
\Psi_{0^{ij}}  &\!=& C_{abcd}\; k^a\, m_i^b\, k^c\, m_j^d \commae \nonumber \\
\Psi_{1^{ijk}} &\!=& C_{abcd}\; k^a\, m_i^b\, m_j^c\, m_k^d  \commae \hspace{2mm}   \Psi_{1T^{i}}= C_{abcd}\; k^a\, l^b\, k^c\, m_i^d \commae\nonumber \\
\Psi_{2^{ijkl}}&\!=& C_{abcd}\; m_i^a\, m_j^b\, m_k^c\, m_l^d \commae \hspace{2mm}  \Psi_{2S}= C_{abcd}\; k^a\, l^b\, l^c\, k^d \commae\nonumber \\
\Psi_{2^{ij}}  &\!=& C_{abcd}\; k^a\, l^b\, m_i^c\, m_j^d \commae \hspace{3mm}      \Psi_{2T^{ij}}= C_{abcd}\; k^a\, m_i^b\, l^c\, m_j^d \commae \nonumber \label{defPsiCoef}\\
\Psi_{3^{ijk}} &\!=& C_{abcd}\; l^a\, m_i^b\, m_j^c\, m_k^d \commae \hspace{3mm}    \Psi_{3T^{i}}= C_{abcd}\; l^a\, k^b\, l^c\, m_i^d \commae\nonumber\\
\Psi_{4^{ij}}  &\!\!=& C_{abcd}\; l^a\, m_i^b\, l^c\, m_j^d \commae
\end{eqnarray}
where ${\,i,j,k,l=2,\ldots,D-1\,}$. All other frame components can be obtained using the symmetries of the Weyl tensor. The scalars in the left column are independent, up to the obvious constraints
\begin{eqnarray}
\Psi_{0^{[ij]}} &\!=& 0 \commae \hspace{17.7mm}  \Psi_{0^{k}}{}^{_k} = 0 \commae\nonumber \\
\Psi_{1^{i(jk)}}&\!=& 0 \commae \hspace{16.5mm}  \Psi_{1^{[ijk]}} = 0 \commae\nonumber \\
\Psi_{2^{ijkl}} &\!=& \Psi_{2^{klij}} \commae \hspace{10.5mm}  \Psi_{2^{(ij)}} = 0 \commae \nonumber \\
\Psi_{2^{(ij)kl}} &\!=& \Psi_{2^{ij(kl)}}=\Psi_{2^{i[jkl]}}=0 \commae \label{psi4sym}\\
\Psi_{3^{i(jk)}}&\!=& 0 \commae \hspace{16.5mm} \Psi_{3^{[ijk]}} = 0 \commae\nonumber \\
\Psi_{4^{[ij]}} &\!=& 0 \commae \hspace{18mm} \Psi_{4^{k}}{}^{_k} = 0 \commae\nonumber
\end{eqnarray}
while those in the right column of \eqref{defPsiCoef} are not independent because they can be expressed as the contractions (hence the symbol ``$T$'' which indicates ``tracing'')
\begin{eqnarray}
\Psi_{1T^i}   &\!=& \Psi_{1^{k}}{}^{_k}{}_{^i} \commae\nonumber \\
\Psi_{2S}     &\!=& \Psi_{2T^{k}}{}^{_k}={\textstyle\frac{1}{2}}\Psi_{2^{kl}}{}^{_{kl}} \commae\label{psi4symT}\\
\Psi_{2T^{ij}}&\!=& {\textstyle\frac{1}{2}}(\Psi_{2^{ikj}}{}^{_k}+\Psi_{2^{ij}}) \nonumber \\
\hbox{where}&& \Psi_{2T^{(ij)}}={\textstyle\frac{1}{2}}\Psi_{2^{ikj}}{}^{_k}\commae \quad \Psi_{2T^{[ij]}}={\textstyle\frac{1}{2}}\Psi_{2^{ij}}\commae\nonumber\\
\Psi_{3T^i}   &\!=& \Psi_{3^{k}}{}^{_k}{}_{^i} \period\nonumber
\end{eqnarray}
In the case ${D=4}$, these Weyl tensor components in the null tetrad reduce to the standard Newman--Penrose \cite{NewmanPenrose:1962,PenroseRindler:book} complex scalars $\Psi_A$. Explicit expressions are given in appendix~\ref{appendixA}.

Using relations ${\,\bolde_{(0)}=\ssqrt(\boldk+\boldl)\,}$, ${\,\bolde_{(1)}=\ssqrt(\boldk-\boldl)\,}$ and the definition \eqref{defPsiCoef}, a straightforward calculation then leads to the following expressions for the components $C_{(\rm{i})(0)(0)(\rm{j})}$ of the Weyl tensor which appear in equation \eqref{InvGeoDevExpl}:
\begin{eqnarray}
C_{(1)(0)(0)(1)} &=& \Psi_{2S} \commae \nonumber \\
C_{(1)(0)(0)(j)} &=& \frac{1}{\sqrt{2}}\,(\,\Psi_{1T^j}-\Psi_{3T^j}) \commae \nonumber \\
C_{(i)(0)(0)(1)} &=& \frac{1}{\sqrt{2}}\,(\,\Psi_{1T^i}-\Psi_{3T^i}) \commae \label{OrthoNullWeylRelations} \\
C_{(i)(0)(0)(j)} &=& -\frac{1}{2}\,(\,\Psi_{0^{ij}}+\Psi_{4^{ij}}) - \Psi_{2T^{(ij)}} \commae \nonumber
\end{eqnarray}
where ${\,i,j=2,\ldots,D-1\,}$ label the spatial directions \emph{orthogonal to} the \emph{privileged spatial direction} of ${\bolde_{(1)}}$.
\begin{widetext}
Putting this into \eqref{InvGeoDevExpl} we obtain the final invariant and fully general form of the equation of geodesic deviation:
\begin{eqnarray}
\ddot{Z}^{(1)}&=&  \frac{2\Lambda }{(D-1)(D-2)}\,Z^{(1)} + \Psi_{2S}\,Z^{(1)}
       + \frac{1}{\sqrt{2}}\,(\,\Psi_{1T^j}-\Psi_{3T^j})\,Z^{(j)} \nonumber\\
      &&  +\frac{8\pi}{D-2}\left[\,T_{(1)(1)} \,Z^{(1)}+T_{(1)(j)} \,Z^{(j)}-\Big(T_{(0)(0)}+\frac{2}{D-1}\,T\Big)\, Z^{(1)}\,\right],\label{InvGeoDevFinal1}\\
\ddot{Z}^{(i)}&=&  \frac{2\Lambda }{(D-1)(D-2)}\,Z^{(i)} - \Psi_{2T^{(ij)}}\,Z^{(j)}
       + \frac{1}{\sqrt{2}}\,(\,\Psi_{1T^i}-\Psi_{3T^i})\,Z^{(1)} -\frac{1}{2}\,(\,\Psi_{0^{ij}}+\Psi_{4^{ij}})\,Z^{(j)} \nonumber\\
      &&  +\frac{8\pi}{D-2}\left[\,T_{(i)(1)} \,Z^{(1)}+T_{(i)(j)} \,Z^{(j)}-\Big(T_{(0)(0)}+\frac{2}{D-1}\,T\Big)\, Z^{(i)}\,\right].\label{InvGeoDevFinali}
\end{eqnarray}
This completely describes relative motion of nearby free test particles in \emph{any spacetime} of an \emph{arbitrary dimension} $D$. In the next section we will discuss the specific effects given by particular scalars which represent the contributions from various components of the gravitational and matter fields.
\end{widetext}

Finally, we remark that our notation which uses $\Psi_{A^{...}}$ in any dimension is simply related to the notations employed, e.g., in  \cite{ColeyMilsonPravdaPravdova:2004,Coley:2008}, in \cite{PravdaPravdovaColeyMilson:2004,PraPraOrt07} and recently in \cite{DurPraPraReall10}. The identifications for the components present in the invariant form of the equation of geodesic deviation are summarized in table~\ref{notationcomp}. More details are given in appendix~\ref{appendixB}, in particular see expressions \eqref{ComparNotation}, \eqref{ComparNotationOther} and \eqref{ComparNotationGHP}.

\begin{table}[h]
\caption{\label{notationcomp}Different equivalent notations used in the literature for the Weyl scalars which occur in the equations of geodesic deviation \eqref{InvGeoDevFinal1}, \eqref{InvGeoDevFinali}.}
\begin{ruledtabular}
\begin{tabular}{lccc}
 & refs.~\cite{ColeyMilsonPravdaPravdova:2004,Coley:2008}  & refs.~\cite{PravdaPravdovaColeyMilson:2004,PraPraOrt07} & ref.~\cite{DurPraPraReall10}  \\
\hline
   $\Psi_{2S}$      & $-C_{0101}$ & $-\Phi$       & $-\Phi$        \\
   $\Psi_{2T^{ij}}$ & $-C_{0i1j}$ & $-\Phi_{ij}$  & $-\Phi_{ij}$   \\
   $\Psi_{1T^{j}}$  & $-C_{010j}$ &               & $-\Psi_j$      \\
   $\Psi_{3T^{j}}$  & $ C_{101j}$ & $\Psi_j$      & $\Psi'_j$      \\
   $\Psi_{0^{ij}}$  & $ C_{0i0j}$ &               & $\Omega_{ij}$  \\
   $\Psi_{4^{ij}}$  & $ C_{1i1j}$ & $2\,\Psi_{ij}$& $\Omega'_{ij}$ \\
\end{tabular}
\end{ruledtabular}
\end{table}


\section{\label{sc:effectscanon}Effect of canonical components of a gravitational field on test particles}
Let us consider a set of freely falling test particles, initially at rest relative to each other, which form, e.g., a small (hyper-)sphere. In any curved spacetime, such a configuration undergoes tidal deformations which can be deduced from the accelerations measured by the fiducial observer attached to the reference test particle in the center. The resulting relative motion represents the effect of a given gravitational field, whose specific structure is explicitly characterized by the system \eqref{InvGeoDevFinal1},~\eqref{InvGeoDevFinali}.

First concentrating on the \emph{vacuum case}, i.e. ${T_{ab}=0}$, the system of equations describing purely gravitational interaction simplifies considerably to
\begin{eqnarray}
\ddot{Z}^{(1)}&=&  \frac{2\Lambda }{(D-1)(D-2)}\,Z^{(1)} + \Psi_{2S}\,Z^{(1)} \label{InvGeoDevFinalVac1}\\
       && + \frac{1}{\sqrt{2}}\,(\,\Psi_{1T^j}-\Psi_{3T^j})\,Z^{(j)}\commae \nonumber\\
\ddot{Z}^{(i)}&=&  \frac{2\Lambda }{(D-1)(D-2)}\,Z^{(i)} - \Psi_{2T^{(ij)}}\,Z^{(j)} \label{InvGeoDevFinalVaci}\\
       && + \frac{1}{\sqrt{2}}\,(\,\Psi_{1T^i}-\Psi_{3T^i})\,Z^{(1)}  -\frac{1}{2}\,(\,\Psi_{0^{ij}}+\Psi_{4^{ij}})\,Z^{(j)} \period \nonumber
\end{eqnarray}
The overall effect of the gravitational field on test particles is thus naturally decomposed into clearly identified components proportional to the cosmological constant $\Lambda$ and the Weyl scalars $\Psi_{A^{...}}$. Of course, for algebraically special spacetimes some (or many) of these coefficients \emph{vanish completely}, and even in algebraically general cases specific \emph{numerical values} of the scalars  $\Psi_{A^{...}}$ can distinguish the dominant terms from those that are negligible. Let us now briefly describe the character of each term separately, including its physical interpretation.

\begin{itemize}
\item $\Lambda\,$: \emph{isotropic influence of the cosmological background}

The presence of the cosmological constant $\Lambda$ is encoded in the term
\begin{equation}\label{Lambda}
\left(
  \begin{array}{c}
    \ddot{Z}^{(1)} \\
    \ddot{Z}^{(i)} \\
  \end{array}
\right)
 = \frac{2\Lambda}{(D-1)(D-2)}
\left(
  \begin{array}{cc}
    1 & 0 \\
    0 & \delta_{ij} \\
  \end{array}
\right) \left(
  \begin{array}{c}
    Z^{(1)} \\
    Z^{(j)} \\
  \end{array}
\right),
\end{equation}
which can be written in a unified way ${ \ddot{Z}^{(\rm{i})}=  \frac{2\Lambda }{(D-1)(D-2)}\,Z^{(\rm{i})} }$ for \emph{all} spatial components ${{\rm{i}}=1,2,\ldots, D-1}$. In parallelly propagated frames, this yields the following explicit solutions
\begin{eqnarray}
\Lambda=0:\quad   && Z^{(\rm{i})} = A_{\rm{i}}\,\tau+B_{\rm{i}} \commae\nonumber \\
\Lambda>0:\quad   && Z^{(\rm{i})} ={\textstyle
     A_{\rm{i}}\cosh\left[\sqrt{\frac{2\Lambda}{(D-1)(D-2)}}\,\tau \right]} \nonumber\\
     &&\hspace{8mm}  {\textstyle+B_{\rm{i}}\sinh\left[\sqrt{\frac{2\Lambda}{(D-1)(D-2)}}\,\tau \right]},\nonumber\\
\Lambda<0:\quad   && Z^{(\rm{i})} ={\textstyle
     A_{\rm{i}}\cos\left[\sqrt{\frac{2\,|\Lambda|}{(D-1)(D-2)}}\,\tau \right]} \nonumber\\
     &&\hspace{8mm}  {\textstyle+B_{\rm{i}}\sin\left[\sqrt{\frac{2\,|\Lambda|}{(D-1)(D-2)}}\,\tau \right]},\nonumber
\end{eqnarray}
where ${A_{\rm{i}},B_{\rm{i}}}$ are constants of integration. These are characteristic relative motions of test particles in spacetimes of constant curvature, namely Minkowski space, de~Sitter space and anti-de~Sitter space, respectively, as derived by Synge \cite{Synge:1934,SyngeSchild:book1949}.

\item $\Psi_{4^{ij}}\,$: \emph{transverse gravitational wave propagating in the direction ${+\bolde_{(1)}}$}

This part of a gravitational field influences the test particles as
\begin{equation}\label{Psi4e}
\left(
  \begin{array}{c}
    \ddot{Z}^{(1)} \\
    \ddot{Z}^{(i)} \\
  \end{array}
\right)
 = -\frac{1}{2}
\left(
  \begin{array}{cc}
    0 & 0 \\
    0 & \Psi_{4^{ij}} \\
  \end{array}
\right) \left(
  \begin{array}{c}
    Z^{(1)} \\
    Z^{(j)} \\
  \end{array}
\right).
\end{equation}
Obviously, this is a purely transverse effect because there is no acceleration in the privileged spatial direction $\bolde_{(1)}$. The set of scalars ${\Psi_{4^{ij}}}$ forms a symmetric (${\Psi_{4^{ij}}=\Psi_{4^{ji}}}$) and traceless ($\Psi_{4^{k}}{}^{_k}=0$) matrix of dimension ${(D-2)\times (D-2)}$, cf. the last line in \eqref{psi4sym}, so that it has ${\frac12 D(D-3)}$ independent components corresponding to polarization modes (see also \cite{Coley:2008,PraPra08,DurPraPraReall10}). In direct analogy with a linearized Einstein gravity in four \cite{MisnerThorneWheeler:book,Wald:book} and higher dimensions \cite{CarDiaLemos03,KrtousPodolsky:2006}, ${\Psi_{4^{ij}}}$ represents the gravitational wave which propagates along the null direction $\boldk$, i.e. in the spatial direction ${+\bolde_{(1)}}$ (in view of relations \eqref{NullFrame} there is ${k_{(1)}\equiv\boldk\cdot\bolde_{(1)}>0}$ while ${k_{(i)}\equiv\boldk\cdot\bolde_{(i)}=0}$ for ${i=2,\ldots,D-1}$). Spacetimes of algebraic type~N (for which only the components  ${\Psi_{4^{ij}} \equiv  C_{1i1j}}$ are nonvanishing \cite{ColeyMilsonPravdaPravdova:2004,Coley:2008}) can thus be interpreted as exact gravitational waves in any dimension ${D \ge 4}$.

\item $\Psi_{3T^{i}}\,$: \emph{longitudinal component of a gravitational field with respect to ${+\bolde_{(1)}}$}

Such terms cause longitudinal deformations of a set of test particles given by
\begin{equation}\label{Psi3e}
\left(
  \begin{array}{c}
    \ddot{Z}^{(1)} \\
    \ddot{Z}^{(i)} \\
  \end{array}
\right)
 = -\frac{1}{\sqrt{2}}
\left(
  \begin{array}{cc}
    0 & \Psi_{3T^j} \\
    \Psi_{3T^i} & 0 \\
  \end{array}
\right) \left(
  \begin{array}{c}
    Z^{(1)} \\
    Z^{(j)} \\
  \end{array}
\right).
\end{equation}
These ${(D-2)}$ scalars ${\Psi_{3T^i}}$, which combine motion in the privileged spatial direction~$\bolde_{(1)}$ with motion in the transverse directions $\bolde_{(i)}$, are also obtained using ${\Psi_{3T^i}\equiv\Psi_{3^{k}}{}^{_k}{}_{^i}}$, where  ${\Psi_{3^{ijk}}=-\Psi_{3^{ikj}}}$ and ${\Psi_{3^{ijk}}+\Psi_{3^{jki}}+\Psi_{3^{kij}}=0}$. Longitudinal effects of this type occur in spacetimes of type~III and in algebraically more general cases.

\item ${\Psi_{2S}, \Psi_{2T^{(ij)}}\,}$: \emph{Newton--Coulomb components of a gravitational field}

The terms
\begin{equation}\label{Psi2e}
\left(
  \begin{array}{c}
    \ddot{Z}^{(1)} \\
    \ddot{Z}^{(i)} \\
  \end{array}
\right)
 =
\left(
  \begin{array}{cc}
    \Psi_{2S} & 0 \\
    0 & -\Psi_{2T^{(ij)}} \\
  \end{array}
\right) \left(
  \begin{array}{c}
    Z^{(1)} \\
    Z^{(j)} \\
  \end{array}
\right)
\end{equation}
give rise to deformations which generalize the classical Newton--Coulomb-type tidal effects in ${D=4}$, namely those in the vicinity of a spherically symmetric static source. Recall that ${\Psi_{2S}=\Psi_{2T^{k}}{}^{_k} }$ (see \eqref{psi4symT} and \eqref{psi4sym} for further relations), so that the ${(D-1)\times (D-1)}$-dimensional matrix in \eqref{Psi2e} is symmetric and traceless. These terms are typically present in type~D spacetimes, for which the notation ${\Psi_{2S} \equiv -\Phi}$ and ${-\Psi_{2T^{(ij)}}\equiv\Phi_{ij}^S}$ is commonly used \cite{PravdaPravdovaColeyMilson:2004,PraPraOrt07,PraPra08,Durkee:2009b,OrtaggioPravdaPravdova:2009,DurPraPraReall10,DurkeeReall:2011}, see \eqref{ComparNotationOther}. As shown in \eqref{D4rel}, the only nonvanishing coefficients of this type in four dimensions are the diagonal elements ${\frac12\Psi_{2S} = \Psi_{2T^{(22)}}= \Psi_{2T^{(33)}} \equiv -\Rea{\Psi_2}}$.

\item $\Psi_{1T^{i}}\,$: \emph{longitudinal component of a gravitational field with respect to ${-\bolde_{(1)}}$}

The corresponding effect on test particles is
\begin{equation}\label{Psi1e}
\left(
  \begin{array}{c}
    \ddot{Z}^{(1)} \\
    \ddot{Z}^{(i)} \\
  \end{array}
\right)
 = \frac{1}{\sqrt{2}}
\left(
  \begin{array}{cc}
    0 & \Psi_{1T^j} \\
    \Psi_{1T^i} & 0 \\
  \end{array}
\right) \left(
  \begin{array}{c}
    Z^{(1)} \\
    Z^{(j)} \\
  \end{array}
\right),
\end{equation}
which is very similar to the acceleration caused by the longitudinal component $\Psi_{3T^{i}}\,$, as described by \eqref{Psi3e}. In fact, it is its counterpart: it follows from the definition \eqref{defPsiCoef} that the scalars ${\Psi_{1T^i}\equiv\Psi_{1^{k}}{}^{_k}{}_{^i}}$ (where ${\Psi_{1^{ijk}}=-\Psi_{1^{ikj}}}$ and ${\Psi_{1^{ijk}}+\Psi_{1^{jki}}+\Psi_{1^{kij}}=0}$) are equivalent to $\Psi_{3T^{i}}\,$ under the interchange ${\boldk \leftrightarrow \boldl}$. Since ${k_{(1)}\equiv\boldk\cdot\bolde_{(1)}>0}$ while ${l_{(1)}\equiv\boldl\cdot\bolde_{(1)}<0}$, the scalars ${\Psi_{1T^i}}$ represent the longitudinal component of the field associated with the spatial direction ${-\bolde_{(1)}}$.

\item $\Psi_{0^{ij}}\,$: \emph{transverse gravitational wave propagating in the direction ${-\bolde_{(1)}}$}

This component of a gravitational field is characterized by
\begin{equation}\label{Psi0e}
\left(
  \begin{array}{c}
    \ddot{Z}^{(1)} \\
    \ddot{Z}^{(i)} \\
  \end{array}
\right)
 = -\frac{1}{2}
\left(
  \begin{array}{cc}
    0 & 0 \\
    0 & \Psi_{0^{ij}} \\
  \end{array}
\right) \left(
  \begin{array}{c}
    Z^{(1)} \\
    Z^{(j)} \\
  \end{array}
\right),
\end{equation}
which is fully equivalent to \eqref{Psi4e} under ${\boldk \leftrightarrow \boldl}$. The scalars ${\Psi_{0^{ij}}}$ (which form a symmetric and traceless ${(D-2)\times (D-2)}$ matrix: ${\Psi_{0^{ij}}=\Psi_{0^{ji}}}$, $\Psi_{0^{k}}{}^{_k}=0$)  thus describe the transverse gravitational wave propagating along the null direction $\boldl$, i.e. in the spatial direction ${-\bolde_{(1)}}$. Superposition of gravitational waves which would propagate in \emph{both} directions simultaneously (that is an ``outgoing'' wave given by ${\Psi_{4^{ij}}}$ and an ``ingoing'' wave given by ${\Psi_{0^{ij}}}$)  can only be present in spacetimes which are of algebraically general type.

\end{itemize}

\section{\label{sc:framechoice}Uniqueness of the interpretation frame and dependence of the field components on the observer}

The canonical components of a gravitational field described in previous section are represented by the real coefficients $\Psi_{A^{...}}$. These are projections of the Weyl tensor onto particular combinations of the null frame ${\{\boldk, \boldl, \boldm_{i} \}}$, as defined in \eqref{defPsiCoef}. They are spacetime scalars and in this sense the above physical interpretation is \emph{invariant}. On the other hand, the values of $\Psi_{A^{...}}$ depend on the choice of the basis vectors of the frame. In this section we will argue that such a dependence corresponds to simple local Lorentz transformations related to the choice of specific observer in a given event, and that the natural interpretation null frame is \emph{essentially unique}.

\begin{figure}[h]
   \includegraphics[scale=0.47]{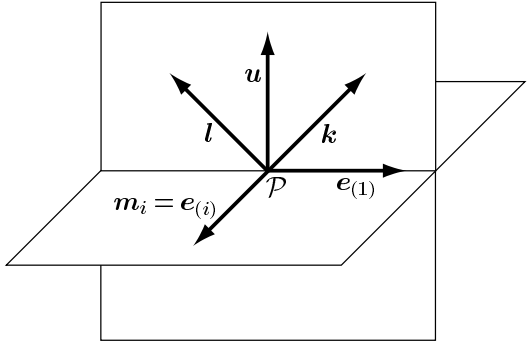}%
      \vspace{-2mm}
   \caption{\label{fig2}%
   Natural choice of the interpretation null frame and the related orthonormal frame \eqref{NullFrame}, \eqref{NullFrNorm}. Up to spatial rotations of ${\boldm_{i}=\bolde_{(i)}}$, they are uniquely given by the velocity vector~$\boldu$ of the observer and the WAND~$\boldk$ at any event ${\cal P}$ of the spacetime.
   }
   \end{figure}

Let us consider an observer attached to the reference (fiducial) test particle moving through some event in the spacetime, such as the point ${\cal P}$ in figure~\ref{fig1}, whose \emph{velocity vector} is $\boldu$. This timelike vector (normalized as ${\boldu\cdot\boldu=-1}$) defines an orthogonal spatial hypersurface of dimension ${D-1}$ spanned by the Cartesian vectors ${\bolde_{(\rm{i})}}$, where ${{\rm{i}}=1,2,\ldots, D-1}$. Assuming the spacetime is of an algebraic type I or more special, it is most natural to associate the corresponding \emph{Weyl aligned null direction} (WAND) with the null vector $\boldk$ of the interpretation reference frame, see figure~\ref{fig2}.

\noindent
The privileged unit vector ${\bolde_{(1)}}$, defining the longitudinal spatial direction, is then uniquely obtained by projecting $\boldk$ onto the spatial subspace orthogonal to $\boldu$. This also fixes the normalization of $\boldk$ (to satisfy the first relation in \eqref{NullFrame} we require ${\boldk\cdot\boldu=-\ssqrt}$). The complementary null vector $\boldl$ of the frame is then also uniquely given via the relation ${\boldl=\sqrt2\,\boldu-\boldk}$. It only remains to choose the transverse spatial vectors ${\bolde_{(2)}, \ldots, \bolde_{(D-1)}}$, i.e. ${\boldm_{i}=\bolde_{(i)}}$. As shown in figure~\ref{fig2}, these must lie in the ${(D-2)}$-dimensional subspace orthogonal both to $\boldu$ and ${\bolde_{(1)}}$, so that ${\boldk\cdot\boldm_{i}=0=\boldl\cdot\boldm_{i}}$ as required by \eqref{NullFrNorm}. Neglecting possible inversions, the only remaining freedom are thus standard spatial rotations represented by the rotation group $SO({D-2})$ which acts on the space spanned by ${\boldm_{i}}$, see the explicit relation \eqref{rotation} presented in appendix~\ref{appendixC}.

For any spacetime of type~N (in which the WAND has maximal alignment order) the null vector $\boldk$ is unique. In spacetimes of other algebraic types (namely $\mbox{III}_i$, $\mbox{II}_i$, $\mbox{I}_i$ and D) different WANDs exist. These can alternatively be used as the vector $\boldk$ of the interpretation null frame ${\{\boldk, \boldl, \boldm_{i} \}}$. Because the distinct WANDs can always be related using the null rotation with fixed ${\boldl}$, as given explicitly by equation \eqref{lfixed} in appendix~\ref{appendixC}, it is straightforward to evaluate the ``new'' values of the Weyl scalars $\Psi_{A^{...}}$ using the expressions \eqref{Weylfixedl}. Notice that the coefficients $\Psi_{4^{ij}}$, which are the amplitudes of transverse gravitational waves propagating along $\boldk$, are invariant under such a change.

Let us now consider \emph{another observer} moving through the \emph{same event} ${\cal P}$ with a \emph{different velocity} $\tilde\boldu$. Locally, this transition is just the Lorentz transformation from the original reference frame ${\{\bolde_a \}}$ to ${\{\tilde\bolde_a \}}$ for which
\begin{equation}\label{generalLT}
\tilde\boldu =\frac{\boldu+\sum_{{\rm{i}}=1}^{D-1}v_{\rm{i}}^{\,}\,\bolde_{(\rm{i})}}{\sqrt{1-\sum_{{\rm{i}}=1}^{D-1} v_{\rm{i}}^2}} \commae
\end{equation}
where ${\,v_1, \ldots, v_{D-1}}$ are components of the spatial velocity of the new observer with respect to the original Cartesian basis ${\bolde_{(\rm{i})}}$. This can be obtained as the combination of a boost in the ${\boldk-\boldl}$ plane  followed by a null rotation with fixed ${\boldk}$, see equations \eqref{boostg} and \eqref{kfixed} in appendix~\ref{appendixC}, if we take the specific parameters
\begin{eqnarray}\label{BL}
B=\frac{\sqrt{1-\sum_{{\rm{i}}=1}^{D-1} v_{\rm{i}}^2}}{1-v_1} \commae\quad L_i=\frac{v_{i}}{\sqrt{1-\sum_{{\rm{i}}=1}^{D-1} v_{\rm{i}}^2}} \commae
\end{eqnarray}
where ${i=2,\ldots,D-1}$:
\begin{eqnarray}
\tilde\boldk    &=&\frac{\sqrt{1-\sum_{{\rm{i}}=1}^{D-1} v_{\rm{i}}^2}}{1-v_1}\,\boldk\commae  \nonumber\\
\tilde\boldl    &=&\frac{1}{\sqrt{1-\sum_{{\rm{i}}=1}^{D-1} v_{\rm{i}}^2}}\bigg[(1-v_1)\,\boldl
      +\sqrt 2\,\sum_{i=2}^{D-1} v_i\,\boldm_{i}   \nonumber\\
        && \hspace{25.0mm} +\frac{\sum_{i=2}^{D-1} v_{i}^2}{1-v_1}\,\boldk\bigg], \\
\tilde\boldm_{i}&=&\boldm_{i}+\sqrt{2}\,\frac{v_i}{1-v_1}\,\boldk \period \nonumber
\end{eqnarray}
Indeed, ${\tilde\boldu \equiv \ssqrt(\tilde\boldk+\tilde\boldl)}$ gives exactly the relation \eqref{generalLT}. The corresponding change of the Weyl scalars $\Psi_{A^{...}}$ can thus be obtained by combining \eqref{Weylboost} with \eqref{Weylfixedk}, which yields
\begin{widetext}
\begin{eqnarray} \label{Weylboostandnullrotation1}
\frac{1}{B^2} \tilde{\Psi}_{0^{ij}} &=&  \Psi_{0^{ij}} \commae \nonumber \\
\frac{1}{B} \,\tilde{\Psi}_{1^{ijk}} &=&  \Psi_{1^{ijk}}-2\sqrt{2}\,\Psi_{0^{i[j}}X_{k]} \commae \nonumber \\
\frac{1}{B} \,\tilde{\Psi}_{1T^{i}} &=&  \Psi_{1T^{i}}+\sqrt{2}\,\Psi_{0^{ij}}X^j \commae \nonumber \\
\quad\tilde{\Psi}_{2^{ijkl}} &=&  \Psi_{2^{ijkl}}-2\sqrt{2}\left(X_{[l}\Psi_{1^{k]ij}}-X_{[i}\Psi_{1^{j]kl}}\right) +4\left(\Psi_{0^{i[k}}X_{l]}X_j+\Psi_{0^{j[l}}X_{k]}X_i\right) \commae \nonumber \\
\quad\tilde{\Psi}_{2S} &=&  \Psi_{2S}-2\sqrt{2}\,\Psi_{1T^{i}}X^i-2\Psi_{0^{ij}}X^iX^j \commae \nonumber \\
\quad\tilde{\Psi}_{2^{ij}} &=&  \Psi_{2^{ij}} +\sqrt{2}\,\Psi_{1^{kij}}X^k -2\sqrt{2}\,\Psi_{1T^{[i}}X_{j]} -4\Psi_{0^{k[i}}X_{j]}X^k \commae \nonumber \\
\quad\tilde{\Psi}_{2T^{ij}} &=&  \Psi_{2T^{ij}}+\sqrt{2}\,\Psi_{1^{ikj}}X^k-\sqrt{2}\,\Psi_{1T^{i}}X_j -2\Psi_{0^{ik}}X^kX_j +\Psi_{0^{ij}}|X|^2 \commae\\
B\,\tilde{\Psi}_{3^{ijk}} &=&  \Psi_{3^{ijk}} +\sqrt{2}\left(\Psi_{2^{lijk}}X^l -\Psi_{2^{jk}}X_i +2X_{[j}\Psi_{2T^{k]i}}\right) \nonumber \\
&& \hspace{5.0mm} + 4\Psi_{1T^{[j}}X_{k]}X_i -2\left(\Psi_{1^{jli}}X_k +\Psi_{1^{ljk}}X_i -\Psi_{1^{kli}}X_j\right)X^l+ \Psi_{1^{ijk}}|X|^2  \nonumber \\
&& \hspace{5.0mm} + 4\sqrt{2}\,\Psi_{0^{l[j}}X_{k]}X_iX^l -2\sqrt{2}\,\Psi_{0^{i[j}}X_{k]}|X|^2 \commae  \nonumber\\
B\,\tilde{\Psi}_{3T^{i}} &=&  \Psi_{3T^{i}} +\sqrt{2}\,\Psi_{2^{ij}}X^j -\sqrt{2}\left(\Psi_{2T^{ki}}X^k+\Psi_{2S}X_i\right) \nonumber \\
&& \hspace{5.0mm} + 2\left(2\Psi_{1T^{j}}X_i -\Psi_{1^{kji}}X^k\right)X^j -\Psi_{1T^{i}}|X|^2  +2\sqrt{2}\,\Psi_{0^{jk}}X^jX^kX_i -\sqrt{2}\,\Psi_{0^{ij}}X^j|X|^2 \commae \nonumber\\
B^2\,\tilde{\Psi}_{4^{ij}} &=&  \Psi_{4^{ij}}+2\sqrt{2}\left(\Psi_{3T^{(i}}X_{j)}-\Psi_{3^{(ij)k}}X^k\right) \nonumber \\
&& \hspace{5.0mm} + 2\Psi_{2^{ikjl}}X^kX^l -4\Psi_{2T^{k(i}}X_{j)}X^k +2\Psi_{2T^{(ij)}}|X|^2 -2\Psi_{2S}X_iX_j -4\Psi_{2^{k(i}}X_{j)}X^k \nonumber \\
&& \hspace{5.0mm} - \, 2\sqrt{2}\,(2\Psi_{1^{kl(i}}X_{j)}X^kX^l +\Psi_{1^{(ij)k}}X^k|X|^2  + \Psi_{1T^{(i}}X_{j)}|X|^2 -2\Psi_{1T^{k}}X^kX_iX_j) \nonumber \\
&& \hspace{5.0mm} + 4\Psi_{0^{kl}}X^kX^lX_iX_j-4\Psi_{0^{k(i}}X_{j)}X^k|X|^2+\Psi_{0^{ij}}|X|^4 \commae \nonumber
\end{eqnarray}
where we denoted
\begin{equation}\label{defX}
X_i\equiv BL_i=\frac{v_i}{1-v_1} \period
\end{equation}
\end{widetext}

In particular, for spacetimes of algebraic type N, which admit a WAND of the maximal alignment order, the only nonvanishing component of the gravitational field is ${\Psi_{4^{ij}}}$ representing the transverse gravitational wave propagating in the spatial direction ${\bolde_{(1)}}$. It immediately follows from \eqref{Weylboostandnullrotation1} and \eqref{BL} that the transition to any other observer results just in a simple rescaling of the gravitational wave amplitudes
\begin{equation}\label{changeN}
\tilde{\Psi}_{4^{ij}} = \frac{(1-v_1)^2}{1-\sum_{{\rm{i}}=1}^{D-1} v_{\rm{i}}^2} \, \Psi_{4^{ij}} \period
\end{equation}
If the new observer moves only in the spatial direction in which the wave propagates, ${v_1>0}$ and ${v_i=0}$ for ${i=2,\ldots, D-1}$. Then ${\tilde{\Psi}_{4^{ij}} = (1-v_1)/(1+v_1) \, \Psi_{4^{ij}}}$ which is smaller than ${\Psi_{4^{ij}}}$. If the observer's velocity approaches the speed of light, ${v_1\to1}$, the amplitudes of the gravitational wave disappear, ${\tilde{\Psi}_{4^{ij}}\to0}$. Contrary, when the observer moves against the wave its amplitudes grow, and for ${v_1\to-1}$ they diverge.

\section{\label{sc:matter}The effect of matter on test particles}

Let us now consider the direct effect of specific forms of matter on relative motion of test particles, as described by the invariant form of the equation of geodesic deviation \eqref{InvGeoDevFinal1}, \eqref{InvGeoDevFinali}. Setting the cosmological constant $\Lambda$ and all components of the Weyl tensor to zero, it reduces to
\begin{eqnarray}
\ddot{Z}^{(1)}&=& \frac{8\pi}{D-2}\bigg[\,T_{(1)(1)} \,Z^{(1)}+T_{(1)(j)} \,Z^{(j)}  \nonumber\\
  && \hspace{9.0mm} -\Big(T_{(0)(0)}+\frac{2}{D-1}\,T\Big)\, Z^{(1)}\,\bigg],\nonumber\\
\ddot{Z}^{(i)}&=& \frac{8\pi}{D-2}\bigg[\,T_{(i)(1)} \,Z^{(1)}\,+T_{(i)(j)} \,Z^{(j)}  \nonumber\\
  && \hspace{9.0mm} -\Big(T_{(0)(0)}+\frac{2}{D-1}\,T\Big)\, Z^{(i)}\,\bigg].\label{InvGeoDevMati}
\end{eqnarray}
It will be illustrative to investigate some important types of matter usually contained in the families of exact solutions of Einstein's equations, namely pure radiation, perfect fluids and electromagnetic fields.

\newpage
\begin{itemize}
\item \emph{pure radiation}

The energy-momentum tensor of a pure radiation field (or ``null dust'') aligned along the null direction $\boldk$ is
\begin{equation}
T_{ab}=\rho\,k_{a}k_{b} \commae
\end{equation}
where $\rho$ is a function representing the radiation density. Its trace vanishes, ${T=0}$, and using \eqref{NullFrame} we derive that the only nonvanishing components of ${T_{ab}}$ in the equation of geodesic deviation are ${T_{(0)(0)}= T_{(1)(1)}=\frac{1}{2}\,\rho}$. Equations \eqref{InvGeoDevMati} thus reduce considerably to
\begin{equation}\label{pure}
\left(
  \begin{array}{c}
    \ddot{Z}^{(1)} \\
    \ddot{Z}^{(i)} \\
  \end{array}
\right)
 = -\frac{4\pi\,\rho}{D-2}\,
\left(
  \begin{array}{cc}
    0 & 0 \\
    0 & \delta_{ij} \\
  \end{array}
\right) \left(
  \begin{array}{c}
    Z^{(1)} \\
    Z^{(j)} \\
  \end{array}
\right).
\end{equation}
In an arbitrary dimension $D$ there is thus no acceleration in the longitudinal spatial direction $\bolde_{(1)}$. The effects in the transverse subspace are isotropic and (since ${\rho>0}$) they cause the radial contraction which may eventually lead to an exact focusing.

\item \emph{perfect fluid}

For a perfect fluid of energy density $\rho$ and pressure $p$ (which is assumed to be isotropic) the energy-momentum tensor is
\begin{equation}
T_{ab}=(\rho+p)\,u_{a}u_{b} + p\,g_{ab} \period
\end{equation}
Provided the fluid is comoving, its velocity $\boldu$ coincides with the observer's velocity which is the vector $\bolde_{(0)}$ of the orthonormal frame. The trace is ${T=(D-1)p-\rho}$, and the relevant nonvanishing frame components are ${T_{(0)(0)}=\rho}$, ${T_{(1)(1)}=p}$ and ${T_{(i)(j)}=p\,\delta_{ij}}$. The equation of geodesic deviation thus takes the form
\begin{equation}\label{fluid}
\left(
  \begin{array}{c}
    \ddot{Z}^{(1)} \\
    \ddot{Z}^{(i)} \\
  \end{array}
\right)
 = -8\pi \frac{ (D-3)\rho + (D-1)p}{(D-1)(D-2)}
\left(
  \begin{array}{cc}
     1 & 0 \\
    0 & \delta_{ij} \\
  \end{array}
\right) \left(
  \begin{array}{c}
    Z^{(1)} \\
    Z^{(j)} \\
  \end{array}
\right).
\end{equation}
The resulting motion is isotropic, the same in the longitudinal and all transverse spatial directions. For positive $\rho$ and $p$, the fluid matter causes a contraction, such as in the case of dust (${p=0}$), incoherent radiation (${p=\frac{D-3}{D-1}\,\rho}$), or stiff fluid (${p=\rho}$). However, for matter with a negative pressure, the set of test particles may expand. In particular, if the matter is described by the equation of state ${p=-\rho=\hbox{const.}}$, it mimics the cosmological constant ${\Lambda=8\pi \rho}$ since \eqref{fluid} is then completely equivalent to \eqref{Lambda}.

\item \emph{electromagnetic field}

The energy-momentum tensor of an electromagnetic field is given by
\begin{equation}
T_{ab}=\frac{1}{4\pi}\Big(F_{ac}\,F_{b}^{\ c}-\frac{1}{4}\,g_{ab}\,F_{cd}\,F^{cd}\Big),
\end{equation}
so that its trace is ${\,T=\frac{1}{16\pi}(4-D)\,F_{ab}\,F^{ab}}$. The frame components of ${T_{ab}}$ which occur in expressions \eqref{InvGeoDevMati} are
\begin{eqnarray}
T_{(0)(0)} &=&  \frac{1}{4\pi}\Big(F_{(0)c}\,{F_{(0)}}^{c}+\frac{1}{4}F_{ab}\,F^{ab}\Big), \nonumber \\
T_{(1)(1)} &=&  \frac{1}{4\pi}\Big(F_{(1)c}\,{F_{(1)}}^{c}-\frac{1}{4}F_{ab}\,F^{ab}\Big), \nonumber \\
T_{(1)(i)} &=&  \frac{1}{4\pi}F_{(1)c}\,{F_{(i)}}^{c} \commae \\
T_{(i)(j)} &=&  \frac{1}{4\pi}\Big(F_{(i)c}\,{F_{(j)}}^{c}-\frac{1}{4}\,\delta_{ij}\,F_{ab}\,F^{ab}\Big). \nonumber
\end{eqnarray}
In this case the equation of geodesic deviation takes the following more complicated form:
\begin{equation}\label{elmag}
\left(
  \begin{array}{c}
    \ddot{Z}^{(1)} \\
    \ddot{Z}^{(i)} \\
  \end{array}
\right)
 =
\left(
  \begin{array}{cc}
    {\cal T}   & {\cal T}_j \\
    {\cal T}_i & {\cal T}_{ij} \\
  \end{array}
\right) \left(
  \begin{array}{c}
    Z^{(1)} \\
    Z^{(j)} \\
  \end{array}
\right),
\end{equation}
where
\begin{eqnarray}
{\cal T}      &=&  \frac{2}{D-2}\,\Big(F_{(1)c}\,{F_{(1)}}^{c}-F_{(0)c}\,{F_{(0)}}^{c}\Big) \nonumber \\
  && -\frac{3}{(D-1)(D-2)}\,F_{ab}\,F^{ab}\commae \nonumber \\
{\cal T}_i    &=&  \frac{2}{D-2}\,F_{(1)c}\,{F_{(i)}}^{c} \commae \\
{\cal T}_{ij} &=&  \frac{2}{D-2}\,\Big(F_{(i)c}\,{F_{(j)}}^{c}-\delta_{ij}\,F_{(0)c}\,{F_{(0)}}^{c}\Big) \nonumber \\
  && -\frac{3}{(D-1)(D-2)}\,\delta_{ij}\,F_{ab}\,F^{ab} \period \nonumber
\end{eqnarray}
We observe that the clear distinction between the longitudinal and transverse spatial directions is not present, except at very special situations. Some important particular subcases can be easily identified and analyzed, for example a null electromagnetic field for which the invariant vanishes, ${F_{ab}\,F^{ab}=0}$, or purely electric aligned field in the vicinity of static black holes.
\end{itemize}

\section{\label{sc:ppwaves}An explicit example: pp-waves in higher dimensions}
We conclude this paper by demonstrating the usefulness of the above interpretation method on an important family of exact spacetimes, namely the pp-waves. These are defined geometrically as admitting a \emph{covariantly constant null vector} field $\boldk$. Such CCNV spacetimes thus form a special subclass of the Kundt spacetimes because the geodesic congruence generated by $\boldk$ is twist-free, shear-free and non-expanding.

In \cite{PodZof09} we investigated general Kundt spacetimes in higher dimensions, admitting a cosmological constant $\Lambda$ and a Maxwell field aligned with $\boldk$ (which is necessarily a multiple WAND). In natural coordinates the metric of all such pp-waves can be written in the Brinkmann form \cite{Bri25}
\begin{equation} \label{gen_metric pp}
 \dif s^2=g_{ij}\,\dif x^i\dif x^j+2\,e_i\,\dif x^i\dif u-2\,\dif u\,\dif r+ c\,\dif u^2 \commae
\end{equation}
where ${\boldk\propto\boldpartial_r}$ and ${g_{ij}, e_i, c}$ are functions of the transverse spatial coordinates $x^k$ and the null coordinate $u$. The explicit Einstein--Maxwell equations can be found in \cite{PodZof09}, namely equations (115)--(118).

For the metric \eqref{gen_metric pp} the interpretation null frame adapted to a general observer which has the velocity ${\boldu=\dot{r}\,\boldpartial_r+\dot{u}\,\boldpartial_u+\dot{x}^2\,\boldpartial_{x^2} + \ldots +  \dot{x}^{D-1}\boldpartial_{x^{D-1}}}$ is
\begin{eqnarray}\label{frame_pp}
\boldk   &=&  \frac{1}{\sqrt{2}\,\dot{u}}\,\boldpartial_r \commae \nonumber \\
\boldl   &=&   \Big(\sqrt{2}\,\dot{r}-\frac{1}{\sqrt{2}\,\dot{u}}\Big)\,\boldpartial_r + \sqrt{2}\,\dot{u}\,\boldpartial_u \nonumber \\
 &&  + \sqrt{2}\,\dot{x}^2\,\boldpartial_{x^2} + \ldots +  \sqrt{2}\,\dot{x}^{D-1}\boldpartial_{x^{D-1}} \commae \\
\boldm_i &=&  \frac{1}{\dot{u}}(e_k\dot{u}+g_{jk}\,\dot{x}^j)\,m_i^k\,\boldpartial_r  \nonumber \\
 &&  + m_i^2\,\boldpartial_{x^2} + \ldots +  m_i^{D-1}\boldpartial_{x^{D-1}} \commae \nonumber
\end{eqnarray}
where ${\,g_{kl}\,m_i^k \,m_j^l}=\delta_{ij}$, and nontrivial components of the Weyl tensor are
\begin{eqnarray}\label{Weyl_pp_coord}
C_{ruru} &=& -\frac{1}{(D-1)(D-2)}\sss R \commae \nonumber \\
C_{riuj} &=& \frac{1}{D-2}\sss R_{ij}-\frac{1}{(D-1)(D-2)}\sss R\,g_{ij} \commae \nonumber \\
C_{ruui} &=& \frac{1}{D-2}\,R_{ui}-\frac{1}{(D-1)(D-2)}\sss R\,e_{i} \commae \nonumber \\
C_{ijkl} &=& \sss R_{ijkl}-\frac{2}{D-2}(g_{i[k}\sss R_{l]j}-g_{j[k}\sss R_{l]i}) \nonumber \\
     &&\qquad\qquad  +\frac{2}{(D-1)(D-2)} \,\sss R\,g_{i[k}\,g_{l]j} \commae \\
C_{uijk} &=& R_{uijk}-\frac{2}{D-2}(e_{[j}\sss R_{k]i}-g_{i[j}\,R_{k]u}) \nonumber \\
     &&\qquad\qquad  +\frac{2}{(D-1)(D-2)} \,\sss R\,e_{[j}\,g_{k]i} \commae \nonumber \\
C_{iuju} &=& R_{iuju}-\frac{1}{D-2}(c \sss R_{ij}-2e_{(i}R_{j)u}+g_{ij}R_{uu}) \nonumber \\
     &&\qquad\qquad  +\frac{1}{(D-1)(D-2)}\,\sss R\,(c\,g_{ij}-e_{i}\,e_{j}) \period \nonumber
\end{eqnarray}
Using definition \eqref{defPsiCoef} we evaluate the Weyl tensor \eqref{Weyl_pp_coord} in the interpretation null frame \eqref{frame_pp}. Lengthy calculation (with some ``miraculous'' cancelations)  gives the following nonvanishing Weyl scalars which enter the equations of geodesic deviation \eqref{InvGeoDevFinal1} and \eqref{InvGeoDevFinali}:
\begin{eqnarray}\label{Weyl_pp_scalars}
\Psi_{2S}      &=& \frac{1}{(D-1)(D-2)} \sss R \commae \nonumber \\
\Psi_{2T^{ij}} &=& \frac{1}{D-2}\sss R_{kl} \,m_{i}^{k}m_{j}^{l}-\frac{1}{(D-1)(D-2)}\sss R\,\delta_{ij} \commae \nonumber \\
\Psi_{3T^i}    &=&  -\frac{\sqrt{2}}{D-2} \big(\sss R_{km}\,\dot{x}^m  + R_{ku}\,\dot{u}\, \big)\, m_i^{k} \commae \\
\Psi_{4^{ij}}  &=& 2\bigg[\Big(\sss R_{kmln}-\frac{1}{D-2}\,g_{kl}\sss R_{mn}\Big)\dot{x}^m\dot{x}^n \nonumber \\
   &&\hspace{1.4mm} +2\Big(R_{kmlu}-\frac{1}{D-2}\,g_{kl}\,R_{mu}\Big)\dot{x}^m\dot{u} \nonumber \\
   &&\hspace{1.4mm} +\Big(R_{kulu}-\frac{1}{D-2}\,g_{kl}\,R_{uu}\Big)\dot{u}^2\bigg]m_{(i}^{k}m_{j)}^{l} \period\nonumber
\end{eqnarray}
This is a general result valid for any pp-wave spacetime because no particular field equations have not yet been imposed.

Notice that ${\Psi_{2T^{ij}}=\Psi_{2T^{(ij)}}}$. Moreover, in accordance with the relations \eqref{psi4symT} and \eqref{psi4sym}, ${\Psi_{2S}=\Psi_{2T^{k}}{}^{_k}}$ and ${\Psi_{4^{k}}{}^{_k} = 0}$ so that any pp-wave is traceless.

The relative tidal motion of nearby test particles in general pp-waves will thus be caused by the combination of the transverse gravitational wave \eqref{Psi4e} propagating along $\boldk$ with amplitude $\Psi_{4^{ij}}$, the longitudinal component \eqref{Psi3e} of the gravitational field with amplitude $\Psi_{3T^i}$, and the Newton--Coulomb contribution \eqref{Psi2e} determined by the scalars $\Psi_{2S}$ and $\Psi_{2T^{ij}}$:
\begin{eqnarray}\label{InvGeoDev1pp}
\ddot{Z}^{(1)}&=&  \ \Psi_{2S}\,Z^{(1)} - {\textstyle \frac{1}{\sqrt{2}}}\,\Psi_{3T^j}\,Z^{(j)}\commae\\
\ddot{Z}^{(i)}&=&  - \Psi_{2T^{(ij)}}\,Z^{(j)} - {\textstyle \frac{1}{\sqrt{2}}}\,\Psi_{3T^i}\,Z^{(1)}  - {\textstyle\frac{1}{2}}\,\Psi_{4^{ij}}\,Z^{(j)} \period \nonumber
\end{eqnarray}
There is also the isotropic background influence \eqref{Lambda} if the cosmological constant $\Lambda$ is present, or the interaction \eqref{elmag} with the electromagnetic field.

The scalars \eqref{Weyl_pp_scalars} which enter \eqref{InvGeoDev1pp} combine \emph{kinematics} (namely the velocity components $\dot{x}^m$, $\dot{u}$ of the observer) with the \emph{specific curvature} of spacetime encoded in the only nonvanishing components of the Riemann and Ricci tensors, namely
\begin{eqnarray}
R_{ijkl} &=& \sss R_{ijkl} \commae \nonumber\\
R_{uijk} &=&
{\textstyle \frac{1}{2}}\left(e_{k,ij}-e_{j,ik}+g_{ij,uk}-g_{ik,uj}\right) \nonumber \\
   && + \sss\,\Gamma^m_{ij}\big({\textstyle \frac{1}{2}}g_{km,u}+e_{[m,k]}\big)-\sss\,\Gamma^m_{ik}\big({\textstyle \frac{1}{2}}g_{jm,u}+e_{[m,j]}\big), \nonumber\\
R_{iuju} &=&  {\textstyle \frac{1}{2}}\left(e_{i,uj}+e_{j,ui}-c_{,ij}-g_{ij,uu}\right)  \\
   && + g^{kl}\big({\textstyle \frac{1}{2}} g_{ik,u}+e_{[k,i]}\big)\big({\textstyle \frac{1}{2}}g_{jl,u}+e_{[l,j]}\big) \nonumber\\
   && -\sss\,\Gamma^k_{ij}\big(e_{k,u}-{\textstyle\frac{1}{2}}c_{,k}\big) , \nonumber
\end{eqnarray}
and
\begin{eqnarray}\label{Ricci_pp}
R_{ij} &=& \sss R_{ij} \commae \nonumber\\
R_{iu} &=& \left[g^{jk}\big({\textstyle \frac{1}{2}}g_{ij,u}+e_{[j,i]}\big)\right]_{,k}  \nonumber\\
   && + \left[g^{jk}\big({\textstyle \frac{1}{2}} g_{ij,u}+e_{[j,i]}\big)\right](\ln\sqrt{g})_{,k} \nonumber \\
   && + g^{jk}g^{lm} \big( g_{im,k}\,e_{[l,j]}-{\textstyle \frac{1}{4}}g_{km,i}\,g_{jl,u}\big) -(\ln\sqrt{g})_{,ui} \commae \nonumber\\
R_{uu} &=& -{\textstyle \frac{1}{2}}(g^{ij}c_{,j})_{,i}-{\textstyle \frac{1}{2}}(g^{ij}c_{,j})(\ln\sqrt{g})_{,i}  \\
   && + (g^{ij}e_{j,u})_{,i}+(g^{ij}e_{j,u})(\ln\sqrt{g})_{,i}+g^{ij}g^{kl}e_{[i,k]}\,e_{j,l} \nonumber\\
   && -{\textstyle \frac{1}{4}}g^{ij}g^{kl}g_{ik,u}\,g_{jl,u}-(\ln\sqrt{g})_{,uu}\commae\nonumber
   \end{eqnarray}
where $\sss R_{ijkl}$ and $\sss R_{ij}$ denote, respectively, the Riemann and Ricci tensors corresponding to the spatial metric $g_{ij}$ only. The Ricci scalar $\sss R $ (equal to $R$) of this transverse $(D-2)$-dimensional Riemannian space enters, in fact, only the Newton--Coulomb scalars $\Psi_{2S}$ and $\Psi_{2T^{ij}}$. Interestingly, these are also \emph{independent of the velocity} of the observer.

There is a big simplification if we restrict ourselves to \emph{vacuum} pp-waves. As shown in \cite{PodZof09}, the absence of an aligned electromagnetic field requires that the cosmological constant $\Lambda$ also vanishes, so that the transverse Riemannian space must be Ricci flat, ${\sss R_{ij} = 0}$. In such a case ${\Psi_{2S}=0=\Psi_{2T^{ij}}}$. Moreover, since ${R_{iu}=0=R_{uu}}$, the Weyl scalar $\Psi_{3T^i}$ also vanishes and the gravitational wave amplitudes reduce to
\begin{eqnarray}\label{Weyl_pp_scalars_vacuum}
\Psi_{4^{ij}}  &=& 2\left[\sss R_{kmln}\,\dot{x}^m\dot{x}^n +2R_{kmlu}\,\dot{x}^m\dot{u}+R_{kulu}\,\dot{u}^2\right]\nonumber \\
   &&\hspace{10mm} \times\, m_{(i}^{k}m_{j)}^{l} \period
\end{eqnarray}

Taking the simplest possibility of a \emph{flat transverse space},
\begin{equation} \label{flatmetric pp}
g_{ij}=\delta_{ij}\commae
\end{equation}
we obtain an important family of exact vacuum plane-fronted gravitational waves (possibly representing an external field of gyratons~\cite{FroFur05,FroIsrZel05}) which propagate in Minkowski space. In fact, these metrics with
\begin{eqnarray}\label{transflatpp}
\sss R_{ijkl} &=& 0 \commae \nonumber\\
R_{uijk} &=& {\textstyle \frac{1}{2}}\left(e_{k,ij}-e_{j,ik}\right), \\
R_{iuju} &=&  {\textstyle \frac{1}{2}}\left(e_{i,uj}+e_{j,ui}-c_{,ij}\right) + \delta^{kl} e_{[k,i]}\,e_{[l,j]}, \nonumber
\end{eqnarray}
belong to the family of VSI spacetimes \cite{ColFusHerPel06}.

If the functions $e_i$ can be globally removed by a suitable coordinate transformation (in the absence of gyratonic sources) the metric reduces to
\begin{equation} \label{spec_metric pp}
 \dif s^2=\delta_{ij}\,\dif x^i\dif x^j-2\,\dif u\,\dif r+ c(x^k,u)\,\dif u^2 \period
\end{equation}
In such a case the spatial vectors of the null frame \eqref{frame_pp} are simply ${ \,\boldm_i = (\dot{x}^i/\dot{u})\,\boldpartial_r  + \boldpartial_{x^i} }$, and the frame is \emph{parallelly transported}. This implies that the physical relative accelerations \eqref{PhysAccel} are, in fact, ordinary time-derivatives of the components of the separation vector, ${\ddot Z^{(\rm{i})} = \frac{\dif^2}{\dif \tau^2}Z^{(\rm{i})} }$. Moreover, ${\dot u=\hbox{const.}}$ along the geodesic since there is ${\Gamma^u_{\alpha\beta}=0}$ for the metric \eqref{spec_metric pp}.

The scalar components of the gravitational field \eqref{Weyl_pp_scalars_vacuum}, \eqref{transflatpp} simplify to
\begin{equation}
\Psi_{4^{ij}} = -\dot{u}^2 \, c_{,ij} \period
\end{equation}
Using \eqref{Ricci_pp}, the only remaining Einstein's vacuum equation ${R_{uu}=0}$ reads ${\Delta \,c\equiv\delta^{ij}\,c_{,ij}=0}$ which explicitly guarantees that the ${(D-2)\times(D-2)}$ symmetric matrix of the wave amplitudes $\Psi_{4^{ij}}$ is traceless. The equations of geodesic deviation thus reduce to
\begin{eqnarray}
\frac{\dif^2{Z}^{(1)}}{\dif \tau^2}&=&  0   \commae \nonumber\\
\frac{\dif^2{Z}^{(i)}}{\dif \tau^2}&=&  {\textstyle \frac{1}{2}}\,\dot{u}^2\,c_{,ij}\,Z^{(j)}  \commae \label{InvGeoDevppis}
\end{eqnarray}
exhibiting the transverse character of the vacuum gravitational pp-waves propagating along ${\bolde_{(1)}}$. In general there are ${\frac12 D(D-3)}$ independent polarization modes corresponding to the same number of free components of the matrix $\Psi_{4^{ij}}$.

In particular, if the metric function $c$ is a \emph{quadratic form} of the transverse spatial coordinates,
\begin{equation}\label{homogc}
c = \sum_{i=2}^{D-1} {\cal A}_i\, (x^i)^2 \commae
\end{equation}
where the constant coefficients ${{\cal A}_i}$ must satisfy
 \begin{equation}\label{homogcA}
\sum_{i=2}^{D-1} {\cal A}_i=0 \commae
\end{equation}
$\Psi_{4^{ij}}$ is a traceless diagonal matrix with \emph{eigenvalues} ${\Psi_{4^{ij}}=-2\,{\cal A}_i\,\dot u^2}$. The amplitudes are constant, i.e., the corresponding gravitational waves are homogeneous. If the test particles are initially at rest [${\dot Z^{(\rm{i})}(\tau=0)=0}$, ${Z^{(\rm{i})}(\tau=0)=Z^{(\rm{i})}_0=\hbox{const.}}$], equations of geodesic deviation \eqref{InvGeoDevppis} for \eqref{homogc} can be explicitly integrated to
\begin{eqnarray}\label{explicsolutpphom}
Z^{(1)}&=& Z^{(1)}_0, \nonumber\\
Z^{(i)}&=& \Bigg\{
\begin{array}{l}
  Z^{(i)}_0\,\cosh\left(\sqrt{{\cal A}_i}\,|\dot u|\,\tau   \right) \quad\hbox{for}\quad {\cal A}_i>0  \commae \\
  Z^{(i)}_0\,\cos\left(\sqrt{{-\cal A}_i}\,|\dot u|\,\tau   \right) \hspace{3mm}\hbox{for}\quad {\cal A}_i<0  \commae \\
  Z^{(i)}_0 \hspace{30mm}\hbox{for}\quad {\cal A}_i=0  \period \\
\end{array}
\end{eqnarray}
Therefore, in the transverse spatial directions ${\bolde_{(i)}}$ with ${{\cal A}_i>0}$ the test particles \emph{recede}, while in those directions with ${{\cal A}_i<0}$ they \emph{focus}. There is also a possibility that ${{\cal A}_i=0}$, in which case there is \emph{no influence of the gravitational wave in the corresponding transverse spatial directions}.

This results in completely new effects which are not allowed in classical ${D=4}$ General Relativity for which ${i=2,3}$ and the constraint \eqref{homogcA} is simply ${{\cal A}_2=-{\cal A}_3}$. Therefore, either a vacuum gravitational pp-wave in 4-dimensional spacetime is absent (${{\cal A}_2=-{\cal A}_3}=0$), or it generates specific particle motions in \emph{both} transverse directions ${\bolde_{(2)}}$ and ${\bolde_{(3)}}$ (focusing in one of them). In higher dimensions, however, the amplitudes are coupled via the $D$-dimensional constraint ${{\cal A}_2=-{\cal A}_3 - \sum_{i=4}^{D-1} {\cal A}_i}$. From the point of view of a detector located on a (1+3)-dimensional \emph{brane} with spatial directions ${\bolde_{(1)}, \bolde_{(2)}, \bolde_{(3)}}$ this would clearly exhibit itself as a \emph{violation of standard TT-property} of gravitational waves (unless ${\sum_{i=4}^{D-1} {\cal A}_i=0}$ which corresponds to a very special subcase). Such an anomalous behavior could possibly serve as a sign of the existence of higher dimensions (see also discussion of a similar effect within the context of linearized 5-dimensional gravitational waves \cite{AlesciMontani:2005}).

It may also happen that ${{\cal A}_k=0}$ for some $k$ (in which case the metric function $c$ given by \eqref{homogc} is independent of the corresponding spatial coordinate $x^k$) and thus there is no effect of the vacuum gravitational pp-wave on test particles in the transverse spatial direction ${\bolde_{(k)}}$. Even the special situations with ${{\cal A}_2=0}$ or ${{\cal A}_3=0}$ are allowed.

\section{Conclusions}

Let us conclude this work by quoting from the classic monograph~\cite{MisnerThorneWheeler:book},~page~35: ``In Einstein's geometric theory of gravity, the equation of geodesic deviation summarizes the entire effect of geometry on matter." This is true not only in standard ${D=4}$ General Relativity, but also in its extension to any higher number of dimensions. Indeed, we have explicitly demonstrated that the geodesic deviation equation, expressed in a suitable reference frame adapted to the observer's geodesic and to the specific algebraic structure of a given spacetime, can be used as a useful tool for analysing and understanding the specific effects of the gravitational field in an arbitrary dimension.

In particular, we derived the general canonical decomposition \eqref{InvGeoDevFinal1}, \eqref{InvGeoDevFinali} of relative accelerations of nearby test particles freely falling in any spacetime. The gravitational contributions, identified and described in section~\ref{sc:effectscanon}, consist of the isotropic background influence \eqref{Lambda} of the cosmological constant $\Lambda$, transverse gravitational waves \eqref{Psi4e}, \eqref{Psi0e}, complementary longitudinal effects \eqref{Psi3e}, \eqref{Psi1e}, and the Newton--Coulomb component \eqref{Psi2e} of the gravitational field. The matter contributions were discussed in section~\ref{sc:matter}, namely the influence of a pure radiation field (null dust) \eqref{pure},  perfect fluid \eqref{fluid}, and generic electromagnetic field \eqref{elmag}.

In the final section~\ref{sc:ppwaves} we also exemplified these results on an important family of exact pp-waves in higher dimensions (admitting a covariantly constant null vector field $\boldk$). Their nontrivial amplitudes are given by expressions \eqref{Weyl_pp_scalars}. The vacuum VSI subclass of such Kundt spacetimes represents purely transverse gravitational waves propagating along the WAND $\boldk$ (in general associated with gyratonic sources). These exact gravitational waves have amplitudes $\Psi_{4^{ij}}$ determined by equations \eqref{Weyl_pp_scalars_vacuum}, \eqref{transflatpp}, which form a ${(D-2)\times(D-2)}$ symmetric traceless matrix. Its ${\frac12 D(D-3)}$ components characterize the independent polarization modes. Explicit solution of the invariant equation of geodesic deviation for the metric function \eqref{homogc} is given in \eqref{explicsolutpphom}. Due to coupling between the eigenvalues of $\Psi_{4^{ij}}$, such higher dimensional gravitational waves could possibly be identified observationally in (1+3)-dimensional brane as a violation of standard TT-property.

We hope that the presented general method of interpreting exact spacetimes, based of the study of geodesic deviation, will help to elucidate the physical and geometrical properties of various explicit solutions of Einstein's equations in an arbitrary dimension.

\begin{acknowledgments}
The work of J.~P. was supported by the grant GA\v{C}R~P203/12/0118 and by the project MSM0021620860. R.~\v{S}. was supported by the grants GA\v{C}R~205/09/H033, SVV-263301 and GAUK~259018.
\end{acknowledgments}

\appendix

\section{\label{appendixA}Relation to complex notation in ${D=4}$}

In standard ${D=4}$ General Relativity it is usual --- instead of the real null frame ${\{\boldk, \boldl, \boldm_{2},\boldm_{3}\}}$ --- to introduce a \emph{complex null tetrad} ${\{\boldk, \boldl, \boldm,\bboldm\}}$ and to parametrize the Weyl tensor by the corresponding five \emph{complex} components. These Newman--Penrose scalar quantities $\Psi_A$, first defined in \cite{NewmanPenrose:1962}, are closely related to the real quantities introduced in our text. Here we present a dictionary relating these two notations. In ${D=4}$ the transverse spatial index ${i}$ runs only over two values ${2,3}$ and we can combine the real vectors ${\boldm_{i}}$ into the complex vectors
\begin{equation}\label{defm4D}
\boldm\equiv{\textstyle\frac1{\sqrt2}}\,(\boldm_{2}-\ci\,\boldm_{3})\,,\qquad \bboldm\equiv{\textstyle\frac1{\sqrt2}}\,(\boldm_{2}+\ci\,\boldm_{3})\period
\end{equation}
Any real spatial vector $\boldV$ spanned on ${\boldm_{2},\boldm_{3}}$ can be parametrized by a complex number $V$ via
the relation
\begin{equation}
\boldV=V^2\boldm_{2}+V^3\boldm_{3}=\ssqrt\,(\bar V\,\boldm+V\,\bboldm)\commae
\end{equation}
so that ${V=V^2-\ci\,V^3}$ and ${|V|^2\equiv(V^2)^2+(V^3)^2=V\bar{V}}$.

In four dimensions there are only \emph{two real independent components} of the Weyl tensor \emph{for each boost weight}, namely
\begin{eqnarray}
&& \Psi_{0^{22}}  =-\Psi_{0^{33}}\,, \qquad  \Psi_{0^{23}}=\Psi_{0^{32}}\commae\nonumber\\
&& \Psi_{1T^{2}}  =\Psi_{1^{332}}=-\Psi_{1^{323}}\,,\qquad  \Psi_{1T^{3}}=\Psi_{1^{223}}=-\Psi_{1^{232}}\commae\nonumber\\
&& \Psi_{2^{2323}}=\Psi_{2^{3232}}=-\Psi_{2^{3223}}=-\Psi_{2^{2332}}=2\,\Psi_{2T^{22}} \nonumber\\
&& \qquad =2\,\Psi_{2T^{33}}=\Psi_{2S}\commae  \nonumber\\
&& \Psi_{2^{23}}  =-\Psi_{2^{32}}=2\,\Psi_{2T^{23}}=-2\,\Psi_{2T^{32}} \commae \nonumber\\
&& \Psi_{3T^{2}}=\Psi_{3^{332}}=-\Psi_{3^{323}}\,,\qquad  \Psi_{3T^{3}}=\Psi_{3^{223}}=-\Psi_{3^{232}}\commae\nonumber \nonumber\\
&& \Psi_{4^{22}}  =-\Psi_{4^{33}}\,, \qquad \Psi_{4^{23}}=\Psi_{4^{32}}\period
\end{eqnarray}
These can be combined into five complex NP components \cite{Stephanietal:book,GriffithPodolsky:book} defined by
\begin{eqnarray}
  \Psi_{0} &=&  C_{abcd}\, k^a \,m^b \,k^c \,m^d \commae\nonumber\\
  \Psi_{1} &=&  C_{abcd}\, k^a \,l^b \,k^c \,m^d \commae\nonumber\\
  \Psi_{2} &=&  C_{abcd}\, k^a \,m^b \,\bar{m}^c \,l^d \commae\\
  \Psi_{3} &=&  C_{abcd}\, l^a \,k^b \,l^c \,\bar{m}^d \commae\nonumber\\
  \Psi_{4} &=&  C_{abcd}\, l^a \,\bar{m}^b \,l^c \,\bar{m}^d \commae\nonumber
\end{eqnarray}
as
\begin{eqnarray}\label{Psi2D4}
  \Psi_{0} &=&  \Psi_{0^{22}}-\ci\,\Psi_{0^{23}} \commae\nonumber\\
  \Psi_{1} &=&  \ssqrt (\Psi_{1T^{2}}-\ci\,\Psi_{1T^{3}}) \commae\nonumber\\
  \Psi_{2} &=&  -{\textstyle\frac12}(\Psi_{2^{2323}}+\ci\,\Psi_{2^{23}}) \commae\\
  \Psi_{3} &=&  \ssqrt (\Psi_{3T^{2}}+\ci\,\Psi_{3T^{3}}) \commae\nonumber\\
  \Psi_{4} &=&  \Psi_{4^{22}}+\ci\,\Psi_{4^{23}} \period\nonumber
\end{eqnarray}
Notice the differences with respect to the notation used in \cite{KrtousPodolsky:2006}: here we have re-labeled all transverse spatial indices as ${i\to i+1}$ to achieve that the privileged spatial direction is denoted as ${\boldm_{1}=\bolde_{(1)}}$, and the scalars ${\Psi_{2S},\Psi_{2T^{ij}}}$ are defined in \eqref{defPsiCoef} without the unnecessary factor 2. (Also, there is a missing factor $\frac{1}{2}$ in equation (A.7c) in \cite{KrtousPodolsky:2006}.)  Inversely, we obtain
\begin{eqnarray}\label{D4rel}
&& \Psi_{0^{22}}= -\Psi_{0^{33}} =  \Rea{\Psi_0} \,, \qquad \Psi_{0^{23}}=\Psi_{0^{32}} = -\Ima{\Psi_0} \commae\nonumber \\
&& \Psi_{1T^{2}}=  \sqrt{2}\,\Rea{\Psi_1} \,, \hspace{16.2mm} \Psi_{1T^{3}}= -\sqrt{2}\,\Ima{\Psi_1} \commae\nonumber \\
&& \Psi_{2S}    = 2\,\Psi_{2T^{22}}= 2\,\Psi_{2T^{33}}=\Psi_{2^{2323}}= -2\Rea{\Psi_2} \,,\nonumber \\
&& \Psi_{2^{23}}= -2\Ima{\Psi_2}\commae\\
&& \Psi_{3T^{2}}=  \sqrt{2}\,\Rea{\Psi_3} \,, \hspace{16.2mm}\Psi_{3T^{3}}=  \sqrt{2}\,\Ima{\Psi_3} \commae\nonumber \\
&& \Psi_{4^{22}}= -\Psi_{4^{33}} =  \Rea{\Psi_4} \,, \qquad \Psi_{4^{23}}=\Psi_{4^{32}} =  \Ima{\Psi_4} \period \nonumber
\end{eqnarray}
According to \eqref{OrthoNullWeylRelations}, the orthonormal components $C_{(\rm{i})(0)(0)(\rm{j})}$ of the Weyl tensor are
\begin{eqnarray}
C_{(1)(0)(0)(1)} &=& -2\Rea{\Psi_2} \commae \nonumber \\
C_{(1)(0)(0)(2)} &=& +\Rea{\Psi_1}-\Rea{\Psi_3} \commae \nonumber \\
C_{(1)(0)(0)(3)} &=& -\Ima{\Psi_1}-\Ima{\Psi_3} \commae \nonumber \\
C_{(2)(0)(0)(2)} &=& \Rea{\Psi_2}-\frac{1}{2}\Rea{\Psi_0}-\frac{1}{2}\Rea{\Psi_4} \commae \\
C_{(3)(0)(0)(3)} &=& \Rea{\Psi_2}+\frac{1}{2}\Rea{\Psi_0}+\frac{1}{2}\Rea{\Psi_4} \commae \nonumber \\
C_{(2)(0)(0)(3)} &=& \frac{1}{2}\Ima{\Psi_0}-\frac{1}{2}\Ima{\Psi_4} \period \nonumber
\end{eqnarray}
Explicit equations of geodesic deviation \eqref{InvGeoDevExpl} in ${D=4}$ thus take the form
\begin{widetext}
\begin{eqnarray}
\ddot{Z}^{(1)}&=&  \frac{\Lambda }{3}\,Z^{(1)}  -2\Rea{\Psi_2}\,Z^{(1)}
       + (\Rea{\Psi_1}-\Rea{\Psi_3})\,Z^{(2)}  - (\Ima{\Psi_1}+\Ima{\Psi_3})\,Z^{(3)}  \nonumber \\
      &&  \qquad+4\pi\left[\,T_{(1)(1)} \,Z^{(1)}+T_{(1)(2)} \,Z^{(2)}+T_{(1)(3)} \,Z^{(3)}-\Big(T_{(0)(0)}+\frac{2}{3}\,T\Big)\, Z^{(1)}\,\right],\nonumber\\
\ddot{Z}^{(2)}&=&  \frac{\Lambda }{3}\,Z^{(2)} +\Rea{\Psi_2}\,Z^{(2)} + (\Rea{\Psi_1}-\Rea{\Psi_3})\,Z^{(1)}
          -\frac{1}{2}\,(\Rea{\Psi_0}+\Rea{\Psi_4})\,Z^{(2)} + \frac{1}{2}\,(\Ima{\Psi_0}-\Ima{\Psi_4})\,Z^{(3)} \nonumber\\
      &&  \qquad+4\pi\left[\,T_{(2)(1)} \,Z^{(1)}+T_{(2)(2)} \,Z^{(2)}+T_{(2)(3)} \,Z^{(3)}-\Big(T_{(0)(0)}+\frac{2}{3}\,T\Big)\, Z^{(2)}\,\right],\label{InvGeoDevFinal2D=4}\\
\ddot{Z}^{(3)}&=&  \frac{\Lambda }{3}\,Z^{(3)} +\Rea{\Psi_2}\,Z^{(3)} - (\Ima{\Psi_1}+\Ima{\Psi_3})\,Z^{(1)}
          + \frac{1}{2}\,(\Ima{\Psi_0}-\Ima{\Psi_4})\,Z^{(2)}+\frac{1}{2}\,(\Rea{\Psi_0}+\Rea{\Psi_4})\,Z^{(3)}  \nonumber \\
      &&  \qquad+4\pi\left[\,T_{(3)(1)} \,Z^{(1)}+T_{(3)(2)} \,Z^{(2)}+T_{(3)(3)} \,Z^{(3)}-\Big(T_{(0)(0)}+\frac{2}{3}\,T\Big)\, Z^{(3)}\,\right].\nonumber
\end{eqnarray}
\end{widetext}
This fully agrees with the results presented in our previous work \cite{BicakPodolsky:1999b} (after permuting the indices as ${1\rightarrow 2\rightarrow 3\rightarrow 1}$, and changing the signs of all imaginary parts due to a convention different from \eqref{defm4D}).

\section{\label{appendixB}Relation to other notations used in ${D\ge 4}$}

In the literature on higher dimensional spacetimes it is common to use alternative conventions for the null frame and the corresponding components. In particular, in the fundamental papers on algebraic classification of the Weyl tensor \cite{ColeyMilsonPravdaPravdova:2004,Coley:2008} the null frame ${\{{\mathbf \ell}, {\mathbf n}, {\mathbf m}_{i} \}}$, where ${i=2,\ldots,D-1}$,
\begin{equation}\label{NullFramePravd}
{\mathbf \ell}\equiv{\mathbf m}_{0} \,, \qquad {\mathbf n}\equiv{\mathbf m}_{1} \,, \qquad {\mathbf m}_{2},\ldots,{\mathbf m}_{D-1} \commae
\end{equation}
is employed such that the metric is  ${\,g_{ab}=2\,\ell_{(a}n_{b)}+\delta_{ij}\,m^{i}_am^{j}_b}$, i.e.
\begin{eqnarray}\label{NullFrNormPravd}
&& {\mathbf \ell}\cdot{\mathbf n}=1 \,, \quad {\mathbf m}_{i}\cdot{\mathbf m}_{j}=\delta_{ij}\,, \nonumber\\
&& {\mathbf \ell}\cdot{\mathbf \ell}=0={\mathbf n}\cdot{\mathbf n}\,, \quad {\mathbf \ell}\cdot{\mathbf m}_{i}=0={\mathbf n}\cdot{\mathbf m}_{i}\period
\end{eqnarray}
Following \cite{ColeyMilsonPravdaPravdova:2004,Coley:2008}, the Weyl tensor can be decomposed into the frame components
\begin{eqnarray}\label{DecompWeylPravd}
C_{abcd} \ = & & \hspace{3.0mm}4\,C_{0i0j}\,n_{\{a}m^{i}_{b}n_{c}m^{j}_{d\}} \nonumber \\
&&  + \, 8\,C_{010i}\,n_{\{a}\ell_{b}n_{c}m^{i}_{d\}}+4\,C_{0ijk}\,n_{\{a}m^{i}_{b}m^{j}_{c}m^{k}_{d\}} \nonumber \\
&&  + \, 4\,C_{0101}\,n_{\{a}\ell_{b}n_{c}\ell_{d\}}+8\,C_{0i1j}\,n_{\{a}m^{i}_{b}\ell_{c}m^{j}_{d\}}   \nonumber \\
&&  \hspace{5.0mm}    +4\,C_{01ij}\,n_{\{a}\ell_{b}m^{i}_{c}m^{j}_{d\}} +C_{ijkl}\,m^{i}_{\{a}m^{j}_{b}m^{k}_{c}m^{l}_{d\}} \nonumber \\
&&  + \, 8\,C_{101i}\,\ell_{\{a}n_{b}\ell_{c}m^{i}_{d\}}+4\,C_{1ijk}\,\ell_{\{a}m^{i}_{b}m^{j}_{c}m^{k}_{d\}} \nonumber \\
&&  + \, 4\,C_{1i1j}\,\ell_{\{a}m^{i}_{b}\ell_{c}m^{j}_{d\}} \commae
\end{eqnarray}
where ${T_{\{abcd\}} \equiv \frac{1}{2}\left(T_{[ab][cd]}+T_{[cd][ab]}\right)}$ is a useful notation representing the standard symmetries of the curvature tensor. The terms in the separate lines of \eqref{DecompWeylPravd} are sorted according to their boost weight corresponding to the scaling
\begin{equation}\label{boost}
\tilde{\mathbf \ell}=\lambda\,{\mathbf \ell} \,,\qquad \tilde{\mathbf n}=\lambda^{-1}{\mathbf n} \,, \qquad \tilde{\mathbf m}_i={\mathbf m}_i \period
\end{equation}
Using \eqref{NullFrNormPravd} we immediately infer that the various scalar components in \eqref{DecompWeylPravd} are explicitly given as
\begin{eqnarray}
C_{0i0j}  &\!=& C_{abcd}\; \ell^a\, m_i^b\, \ell^c\, m_j^d \commae \nonumber \\
C_{0ijk}  &\!=& C_{abcd}\; \ell^a\, m_i^b\, m_j^c\, m_k^d  \commae \hspace{2mm}   C_{010i}  =  C_{abcd}\; \ell^a\, n^b\, \ell^c\, m_i^d \commae \nonumber \\
 C_{ijkl} &\!=& C_{abcd}\; m_i^a\, m_j^b\, m_k^c\, m_l^d   \commae \hspace{0mm}   C_{0101} =  C_{abcd}\; \ell^a\, n^b\, \ell^d\, n^c  \commae \nonumber \\
C_{01ij}  &\!=& C_{abcd}\; \ell^a\, n^b\, m_i^c\, m_j^d    \commae \hspace{3.7mm} C_{0i1j}  =  C_{abcd}\; \ell^a\, m_i^b\, n^c\, m_j^d \commae \nonumber\\
 C_{1ijk} &\!=& C_{abcd}\; n^a\, m_i^b\, m_j^c\, m_k^d     \commae \hspace{2mm}   C_{101i} =  C_{abcd}\; n^a\, \ell^b\, n^c\, m_i^d \commae \nonumber\\
 C_{1i1j} &\!=& C_{abcd}\; n^a\, m_i^b\, n^c\, m_j^d       \period \label{defPsiCoefPravd}
\end{eqnarray}
They are subject to a number of mutual relations which follow from the symmetries and from the trace-free property of the Weyl tensor, see \cite{Coley:2008}:
\begin{eqnarray}\label{reltionsPravd}
{C_{0i0}}^i &\!=&  0  \,,\quad C_{010j}={C_{0ij}}^i  \,,\quad C_{0[ijk]}=0  \,,\nonumber\\
 C_{0101} &\!=&  {C_{0i1}}^i \,,\quad C_{i[jkl]}=0\commae\nonumber\\
 C_{0i1j} &\!=& -{\textstyle\frac12} {C_{ikj}}^k+{\textstyle\frac12} C_{01ij} \,,\quad C_{011j}=-{C_{1ij}}^i  \,,\nonumber\\
 C_{1[ijk]} &\!=& 0  \,,\quad {C_{1i1}}^i=0 \period
\end{eqnarray}
Now, by comparing \eqref{NullFrNormPravd} with our definition \eqref{NullFrNorm} it follows that the two null frames \eqref{NullFramePravd} and \eqref{NullFrame} are related simply  as
\begin{equation}\label{ConvRelation}
\boldk\equiv{\mathbf \ell} \,, \qquad \boldl\equiv -{\mathbf n} \,, \qquad \boldm_{i}\equiv {\mathbf m}_{i} \period
\end{equation}
Putting this identification into \eqref{defPsiCoefPravd}, and comparing with \eqref{defPsiCoef}, we observe that
\begin{eqnarray}\label{ComparNotation}
\Psi_{0^{ij}}  &\!\equiv &  C_{0i0j}\commae \nonumber \\
\Psi_{1^{ijk}} &\!\equiv &  C_{0ijk} \,, \hspace{6.5mm}  \Psi_{1T^{i}} \equiv -C_{010i}\commae \nonumber \\
\Psi_{2^{ijkl}}&\!\equiv &  C_{ijkl} \,, \hspace{8mm}  \Psi_{2^{ij}} \equiv -C_{01ij}  \,,  \nonumber \\
\Psi_{2S} &\!\equiv &  -C_{0101}  \,, \hspace{2.5mm} \Psi_{2T^{ij}} \equiv  -C_{0i1j}\,, \nonumber\\
\Psi_{3^{ijk}} &\!\equiv & -C_{1ijk} \,, \hspace{4.2mm}  \Psi_{3T^{i}} \equiv  C_{101i} \commae \nonumber\\
\Psi_{4^{ij}}  &\!\equiv &  C_{1i1j} \period
\end{eqnarray}
Moreover, the relations \eqref{reltionsPravd} are equivalent to the constraints \eqref{psi4sym}, \eqref{psi4symT}.

Also, in \cite{PravdaPravdovaColeyMilson:2004,ColMilPraPra04,ColFusHerPel06,OrtPraPra10} the notation
\begin{equation}\label{PSIS}
\Psi_{ij}\equiv{\textstyle\frac{1}{2}}C_{1i1j} \,, \qquad \Psi_{ijk}\equiv{\textstyle\frac{1}{2}}C_{1kij} \,, \qquad  \Psi_i\equiv C_{101i} \commae
\end{equation}
was introduced and employed, which is useful for studies of type N and type III spacetimes, and
\begin{eqnarray}\label{PHIS}
&& \Phi_{ij}\equiv C_{0i1j} \,, \qquad \Phi_{ij}^A\equiv{\textstyle\frac{1}{2}}C_{01ij} \,, \qquad \Phi_{ij}^S\equiv{\textstyle-\frac{1}{2}}{C_{ikj}}^k \,, \nonumber\\
&& \Phi\equiv C_{0101}={\textstyle-\frac{1}{2}}{C_{ij}}^{ij}\period
\end{eqnarray}
(where ${\Phi_{ij}^A,\Phi_{ij}^S,\Phi}$ denote antisymmetric, symmetric parts of $\Phi_{ij}$ and its trace, respectively) which is convenient for type D spacetimes \cite{PraPraOrt07,PraPra08,Durkee:2009b,OrtaggioPravdaPravdova:2009}. In view of \eqref{ComparNotation} we thus easily identify
\begin{eqnarray}\label{ComparNotationOther}
\Psi_{2^{ijkl}}&\!\equiv &  C_{ijkl} \,, \hspace{7.9mm}  \Psi_{2^{ij}} \equiv -2\,\Phi_{ij}^A  \,,  \nonumber\\
  \Psi_{2S} &\!\equiv &  -\Phi  \,, \hspace{8mm} \Psi_{2T^{ij}} \equiv -\Phi_{ij}\,, \nonumber\\
\Psi_{3^{ijk}} &\!\equiv & -2\,\Psi_{jki} \,, \hspace{2.4mm}  \Psi_{3T^{i}} \equiv  \Psi_i \commae \nonumber\\
\Psi_{4^{ij}}  &\!\equiv &  2\,\Psi_{ij} \period
\end{eqnarray}

Very recently, in the generalisation of the Geroch--Held--Penrose formalism to higher dimensions \cite{DurPraPraReall10}, another convention was suggested, namely
\begin{eqnarray}\label{GHPform}
\Omega_{ij}  &\!\equiv & C_{0i0j} \,, \qquad \Psi_{ijk} \equiv C_{0ijk} \,, \qquad \Psi_i \equiv C_{010i} \commae\nonumber\\
\Omega'_{ij} &\!\equiv & C_{1i1j} \,, \qquad \Psi'_{ijk}\equiv C_{1ijk} \,, \qquad \Psi'_i\equiv C_{101i} \commae\nonumber\\
\qquad \Phi_{ijkl} &\!\equiv & C_{ijkl} \period
\end{eqnarray}
These scalars are straightforwardly related to the quantities used in the present paper:
\begin{eqnarray}\label{ComparNotationGHP}
\Psi_{0^{ij}}  &\!\equiv &  \Omega_{ij} \commae \nonumber \\
\Psi_{1^{ijk}} &\!\equiv &  \Psi_{ijk} \,, \hspace{7.6mm}  \Psi_{1T^{i}} \equiv -\Psi_i\commae \nonumber \\
\Psi_{2^{ijkl}}&\!\equiv &  \Phi_{ijkl} \,, \hspace{7.9mm}  \Psi_{2^{ij}} \equiv -2\,\Phi_{ij}^A  \commae \nonumber \\
     \Psi_{2S} &\!\equiv & -\Phi  \,, \hspace{8.1mm} \Psi_{2T^{ij}} \equiv -\Phi_{ij}\,, \nonumber\\
\Psi_{3^{ijk}} &\!\equiv & -\Psi'_{ijk} \,, \hspace{5.0mm}  \Psi_{3T^{i}} \equiv  \Psi'_i \commae \nonumber\\
\Psi_{4^{ij}}  &\!\equiv &  \Omega'_{ij} \period
\end{eqnarray}

\section{\label{appendixC}Lorentz transformations of the null frame and the changes of $\Psi_{A^{...}}$}

It is well known (see e.g. \cite{ColeyMilsonPravdaPravdova:2004,Coley:2008,KrtousPodolsky:2006}) that general transformations between different null frames can be composed from
the following simple Lorentz transformations:

$\bullet$  \emph{null rotation with ${\boldk}$ fixed} (parametrized by ${D-2}$ real parameters $L^i$):
\begin{equation}\label{kfixed}
  \tilde{\boldk} = \boldk \commae\quad
  \tilde{\boldl} = \boldl + \sqrt2\, L^i \boldm_{i} + |L|^2\, \boldk\commae\quad
  \tilde{\boldm}_{i} = \boldm_{i} + \sqrt2\, L_i\,\boldk\commae
\end{equation}

$\bullet$  \emph{null rotation with ${\boldl}$ fixed} (parametrized by ${D-2}$ real parameters $K^i$):
\begin{equation}\label{lfixed}
  \tilde{\boldk} = \boldk + \sqrt2\, K^i \boldm_{i} + |K|^2\, \boldl\commae\quad
  \tilde{\boldl} = \boldl \commae\quad
  \tilde{\boldm}_{i} = \boldm_{i} + \sqrt2\, K_i\,\boldl\commae
\end{equation}

$\bullet$  \emph{boost in the ${\boldk-\boldl}$ plane} (parametrized by a real number ${B}$):
\begin{equation}\label{boostg}
  \tilde{\boldk} = B\,\boldk \commae\quad
  \tilde{\boldl} = B^{-1}\, \boldl \commae\quad
  \tilde{\boldm}_{i} = \boldm_{i} \commae
\end{equation}

$\bullet$  \emph{spatial rotation in the space of ${\boldm_{i}}$} (parametrized by an orthogonal matrix ${\Phi_i{}^j}$):
\begin{equation}\label{rotation}
  \tilde{\boldk} = \boldk \commae\
  \tilde{\boldl} = \boldl \commae\
  \tilde{\boldm}_{i} = \Phi_i{}^j\,\boldm_{j} \commae\ \text{with}\  \Phi_i{}^j\,\Phi_k{}^l\;\delta_{jl}=\delta_{ik}\period
\end{equation}

Due to \eqref{NullFrNorm}  ${L^i=L_i}$, ${K^i=K_i}$ and we employ a shorthand ${|L|^2\equiv L^i L_i}$, ${|K|^2\equiv K^i K_i}$. Under these Lorentz transformations of the frame, the Weyl scalars change as:

$\bullet$ null rotation with ${\boldk}$ fixed
\begin{widetext}
\begin{eqnarray}\label{Weylfixedk}
\tilde{\Psi}_{0^{ij}}&\!=& \Psi_{0^{ij}} \commae \nonumber \\
\tilde{\Psi}_{1^{ijk}}&\!=& \Psi_{1^{ijk}}-2\sqrt{2}\,\Psi_{0^{i[j}}L_{k]} \commae \nonumber \\
\tilde{\Psi}_{1T^{i}}&\!=& \Psi_{1T^{i}}+\sqrt{2}\,\Psi_{0^{ij}}L^j \commae \nonumber \\
\tilde{\Psi}_{2^{ijkl}}&\!=& \Psi_{2^{ijkl}}-2\sqrt{2}\left(L_{[l}\Psi_{1^{k]ij}}-L_{[i}\Psi_{1^{j]kl}}\right) +4(\Psi_{0^{i[k}}L_{l]}L_j+\Psi_{0^{j[l}}L_{k]}L_i)\commae \nonumber \\
\tilde{\Psi}_{2S}&\!=& \Psi_{2S}-2\sqrt{2}\,\Psi_{1T^{i}}L^i-2\Psi_{0^{ij}}L^iL^j \commae \nonumber \\
\tilde{\Psi}_{2^{ij}}&\!=& \Psi_{2^{ij}} +\sqrt{2}\,\Psi_{1^{kij}}L^k -2\sqrt{2}\,\Psi_{1T^{[i}}L_{j]} -4\Psi_{0^{k[i}}L_{j]}L^k \commae \nonumber \\
\tilde{\Psi}_{2T^{ij}}&\!=& \Psi_{2T^{ij}}+\sqrt{2}\,\Psi_{1^{ikj}}L^k-\sqrt{2}\,\Psi_{1T^{i}}L_j -2\Psi_{0^{ik}}L^kL_j +\Psi_{0^{ij}}|L|^2 \commae\nonumber\\
\tilde{\Psi}_{3^{ijk}}&\!=& \Psi_{3^{ijk}} +\sqrt{2}\left(\Psi_{2^{lijk}}L^l -\Psi_{2^{jk}}L_i +2L_{[j}\Psi_{2T^{k]i}}\right) \nonumber \\
&&  + 4\Psi_{1T^{[j}}L_{k]}L_i -2\left(\Psi_{1^{jli}}L_k +\Psi_{1^{ljk}}L_i -\Psi_{1^{kli}}L_j\right)L^l + \Psi_{1^{ijk}}|L|^2  \nonumber \\
&&  + 4\sqrt{2}\,\Psi_{0^{l[j}}L_{k]}L_iL^l -2\sqrt{2}\,\Psi_{0^{i[j}}L_{k]}|L|^2 \commae \\
\tilde{\Psi}_{3T^{i}}&\!=& \Psi_{3T^{i}} +\sqrt{2}\,\Psi_{2^{ij}}L^j -\sqrt{2}\left(\Psi_{2T^{ki}}L^k+\Psi_{2S}L_i\right) \nonumber \\
&&  + 2\left(2\Psi_{1T^{j}}L_i -\Psi_{1^{kji}}L^k\right)L^j -\Psi_{1T^{i}}|L|^2  + 2\sqrt{2}\,\Psi_{0^{jk}}L^jL^kL_i -\sqrt{2}\,\Psi_{0^{ij}}L^j|L|^2 \commae \nonumber\\
\tilde{\Psi}_{4^{ij}}&\!=& \Psi_{4^{ij}}+2\sqrt{2}\left(\Psi_{3T^{(i}}L_{j)}-\Psi_{3^{(ij)k}}L^k\right) \nonumber \\
&&  + 2\Psi_{2^{ikjl}}L^kL^l -4\Psi_{2T^{k(i}}L_{j)}L^k +2\Psi_{2T^{(ij)}}|L|^2  - 2\Psi_{2S}L_iL_j -4\Psi_{2^{k(i}}L_{j)}L^k \nonumber \\
&&  - 2\sqrt{2}\,(2\Psi_{1^{kl(i}}L_{j)}L^kL^l +\Psi_{1^{(ij)k}}L^k|L|^2  + \Psi_{1T^{(i}}L_{j)}|L|^2 -2\Psi_{1T^{k}}L^kL_iL_j) \nonumber \\
&&  + 4\Psi_{0^{kl}}L^kL^lL_iL_j-4\Psi_{0^{k(i}}L_{j)}L^k|L|^2+\Psi_{0^{ij}}|L|^4 \commae \nonumber
\end{eqnarray}

$\bullet$  null rotation with ${\boldl}$ fixed
\begin{eqnarray}\label{Weylfixedl1}
\tilde{\Psi}_{0^{ij}}&\!=& \Psi_{0^{ij}}+2\sqrt{2}\left(\Psi_{1T^{(i}}K_{j)}-\Psi_{1^{(ij)k}}K^k\right) \nonumber \\
&&  + 2\Psi_{2^{ikjl}}K^kK^l -4K_{(i}\Psi_{2T^{j)k}}K^k+2\Psi_{2T^{(ij)}}|K|^2  -2\Psi_{2S}K_iK_j +4\Psi_{2^{k(i}}K_{j)}K^k \nonumber \\
&&  - 2\sqrt{2}\,(2\Psi_{3^{kl(i}}K_{j)}K^kK^l +\Psi_{3^{(ij)k}}K^k|K|^2 +\Psi_{3T^{(i}}K_{j)}|K|^2 -2\Psi_{3T^{k}}K^kK_iK_j) \nonumber \\
&&  + 4\Psi_{4^{kl}}K^kK^lK_iK_j-4\Psi_{4^{k(i}}K_{j)}K^k|K|^2+\Psi_{4^{ij}}|K|^4 \commae \nonumber \\
\tilde{\Psi}_{1^{ijk}}&\!=& \Psi_{1^{ijk}} +\sqrt{2}\left(\Psi_{2^{lijk}}K^l +\Psi_{2^{jk}}K_i -2\Psi_{2T^{i[j}}K_{k]}\right) \nonumber \\
&&  + 4\Psi_{3T^{[j}}K_{k]}K_i -2\left(\Psi_{3^{jli}}K_k +\Psi_{3^{ljk}}K_i -\Psi_{3^{kli}}K_j\right)K^l + \Psi_{3^{ijk}}|K|^2 \nonumber \\
&&  + 4\sqrt{2}\,\Psi_{4^{l[j}}K_{k]}K_iK^l -2\sqrt{2}\,\Psi_{4^{i[j}}K_{k]}|K|^2 \commae \\
\tilde{\Psi}_{1T^{i}}&\!=& \Psi_{1T^{i}} +\sqrt{2}\,\Psi_{2^{ji}}K^j -\sqrt{2}\left(\Psi_{2T^{ij}}K^j+\Psi_{2S}K_i\right) \nonumber \\
&&  + 2\left(2\Psi_{3T^{j}}K_i -\Psi_{3^{kji}}K^k\right)K^j -\Psi_{3T^{i}}|K|^2  +2\sqrt{2}\,\Psi_{4^{jk}}K^jK^kK_i -\sqrt{2}\,\Psi_{4^{ij}}K^j|K|^2 \commae \nonumber
\end{eqnarray}
\begin{eqnarray}\label{Weylfixedl2}
\tilde{\Psi}_{2^{ijkl}}&\!=& \Psi_{2^{ijkl}}-2\sqrt{2}\left(K_{[l}\Psi_{3^{k]ij}}-K_{[i}\Psi_{3^{j]kl}}\right) +4(\Psi_{4^{i[k}}K_{l]}K_j+\Psi_{4^{j[l}}K_{k]}K_i) \commae \nonumber \\
\tilde{\Psi}_{2S}&\!=& \Psi_{2S}-2\sqrt{2}\,\Psi_{3T^{i}}K^i-2\Psi_{4^{ij}}K^iK^j \commae \nonumber \\
\tilde{\Psi}_{2^{ij}}&\!=& \Psi_{2^{ij}} -\sqrt{2}\,\Psi_{3^{kij}}K^k +2\sqrt{2}\,\Psi_{3T^{[i}}K_{j]} +4\Psi_{4^{k[i}}K_{j]}K^k \commae \nonumber \\
\tilde{\Psi}_{2T^{ij}}&\!=& \Psi_{2T^{ij}}+\sqrt{2}\,\Psi_{3^{jki}}K^k-\sqrt{2}\,\Psi_{3T^{j}}K_i -2\Psi_{4^{jk}}K^kK_i +\Psi_{4^{ij}}|K|^2 \commae \nonumber \\
\tilde{\Psi}_{3^{ijk}}&\!=& \Psi_{3^{ijk}}-2\sqrt{2}\,\Psi_{4^{i[j}}K_{k]} \commae \nonumber \\
\tilde{\Psi}_{3T^{i}}&\!=& \Psi_{3T^{i}}+\sqrt{2}\,\Psi_{4^{ij}}K^j \commae \nonumber \\
\tilde{\Psi}_{4^{ij}}&\!=& \Psi_{4^{ij}} \commae \nonumber
\end{eqnarray}

$\bullet$  boost in the ${\boldk-\boldl}$ plane
\begin{eqnarray}\label{Weylboost}
\tilde{\Psi}_{0^{ij}}  &\!=& B^2\,\Psi_{0^{ij}} \commae \nonumber \\
\tilde{\Psi}_{1^{ijk}} &\!=& B\,\Psi_{1^{ijk}}  \commae \hspace{5.5mm} \tilde{\Psi}_{1T^{i}}=B\,\Psi_{1T^{i}} \commae \nonumber \\
\tilde{\Psi}_{2^{ijkl}}&\!=& \Psi_{2^{ijkl}}  \commae \hspace{9mm} \tilde{\Psi}_{2^{ij}}=\Psi_{2^{ij}} \commae \hspace{6mm}  \tilde{\Psi}_{2S}=\Psi_{2S} \commae \hspace{6mm} \tilde{\Psi}_{2T^{ij}}=\Psi_{2T^{ij}}\commae \\
\tilde{\Psi}_{3^{ijk}} &\!=& B^{-1}\Psi_{3^{ijk}}\commae \hspace{2.4mm} \tilde{\Psi}_{3T^{i}}=B^{-1}\Psi_{3T^{i}}\commae\nonumber \\
\tilde{\Psi}_{4^{ij}} &\!=& B^{-2}\Psi_{4^{ij}} \commae \nonumber
\end{eqnarray}

$\bullet$  spatial rotation in the space of ${\boldm_{i}}$
\begin{eqnarray}\label{Weylspatrotation}
\tilde{\Psi}_{0^{ij}}&\!=& \Phi_{i}^{\ p}\,\Phi_{j}^{\ q}\,\Psi_{0^{pq}} \commae \nonumber \\
\tilde{\Psi}_{1^{ijk}}&\!=& \Phi_{i}^{\ o}\,\Phi_{j}^{\ p}\,\Phi_{k}^{\ q}\,\Psi_{1^{opq}} \commae \hspace{12mm} \tilde{\Psi}_{1T^{i}}=\Phi_{i}^{\ p}\,\Psi_{1T^{p}} \commae \nonumber \\
\tilde{\Psi}_{2^{ijkl}}&\!=& \Phi_{i}^{\ n}\,\Phi_{j}^{\ o}\,\Phi_{k}^{\ p}\,\Phi_{l}^{\ q}\,\Psi_{2^{nopq}} \commae \hspace{5.5mm} \tilde{\Psi}_{2S}=\Psi_{2S} \commae \nonumber \\
 \tilde{\Psi}_{2^{ij}} &\!=& \Phi_{i}^{\ p}\,\Phi_{j}^{\ q}\,\Psi_{2^{pq}} \commae \hspace{17.8mm} \tilde{\Psi}_{2T^{ij}}=\Phi_{i}^{\ p}\,\Phi_{j}^{\ q}\,\Psi_{2T^{pq}} \commae \nonumber \\
\tilde{\Psi}_{3^{ijk}}&\!=& \Phi_{i}^{\ o}\,\Phi_{j}^{\ p}\,\Phi_{k}^{\ q}\,\Psi_{3^{opq}} \commae \hspace{12mm}  \tilde{\Psi}_{3T^{i}}=\Phi_{i}^{\ p}\,\Psi_{3T^{p}} \commae \\
\tilde{\Psi}_{4^{ij}}&\!=& \Phi_{i}^{\ p}\,\Phi_{j}^{\ q}\,\Psi_{4^{pq}} \period \nonumber
\end{eqnarray}
\end{widetext}

\end{document}